\documentclass[sigconf]{acmart}
\theoremstyle{definition}
\newtheorem{definition}{Definition}
\usepackage{subfig}
\usepackage{graphics}
\usepackage{float}
\usepackage{natbib}
\usepackage{graphicx}
\usepackage{caption}
\usepackage{color}
\usepackage{changepage}
\usepackage{graphicx}
\usepackage{amsmath}
\usepackage[linesnumbered,ruled]{algorithm2e}

\AtBeginDocument{%
  }

\setcopyright{none}
\renewcommand\footnotetextcopyrightpermission[1]{} 
\begin{document}

\title{DeepScalper: A Risk-Aware Reinforcement Learning Framework to Capture Fleeting Intraday Trading Opportunities}


\author{Shuo Sun}
\affiliation{%
  \institution{Nanyang Technological University}
  \country{}}

\author{Wanqi Xue}
\affiliation{%
  \institution{Nanyang Technological University}
  \country{}}

\author{Rundong Wang}
\authornote{Corresponding author.}
\affiliation{%
  \institution{Nanyang Technological University}
  \country{}}

\author{Xu He}
\affiliation{%
  \institution{Huawei Noah Ark Lab}
  \country{}}

\author{Junlei Zhu}
\affiliation{%
  \institution{Webank}
  \country{}}

\author{Jian Li}
\affiliation{%
  \institution{Tsinghua University}
  \country{}}

\author{Bo An}
\affiliation{%
  \institution{Nanyang Technological University}
  \country{}}

\renewcommand{\shortauthors}{Sun et al.}

\begin{abstract}
Reinforcement learning (RL) techniques have shown great success in many challenging quantitative trading tasks, such as portfolio management and algorithmic trading. Especially, intraday trading is one of the most profitable and risky tasks because of the intraday behaviors of the financial market that reflect billions of rapidly fluctuating capitals. However, a vast majority of existing RL methods focus on the relatively low frequency trading scenarios (e.g., day-level) and fail to capture the fleeting intraday investment opportunities due to two major challenges: 1) how to effectively train profitable RL agents for intraday investment decision-making, which involves high-dimensional fine-grained action space; 2) how to learn meaningful multi-modality market representation to understand the intraday behaviors of the financial market at tick-level.

Motivated by the efficient workflow of professional human intraday traders, we propose DeepScalper, a deep reinforcement learning framework for intraday trading to tackle the above challenges. Specifically, DeepScalper includes four components: 1) a dueling Q-network with action branching to deal with the large action space of intraday trading for efficient RL optimization; 2) a novel reward function with a hindsight bonus to encourage RL agents making trading decisions with a long-term horizon of the entire trading day; 3) an encoder-decoder architecture to learn multi-modality temporal market embedding, which incorporates both macro-level and micro-level market information; 4) a risk-aware auxiliary task to maintain a striking balance between maximizing profit and minimizing risk. Through extensive experiments on real-world market data spanning over three years on six financial futures (2 stock index and 4 treasury bond), we demonstrate that DeepScalper significantly outperforms many state-of-the-art baselines in terms of four financial criteria. Furthermore, we conduct a series of exploratory and ablative studies to analyze the contributions of each component in DeepScalper.
\end{abstract}


\maketitle

\section{Introduction}
The financial market, an ecosystem that involves over $90$ trillion\footnote{https://data.worldbank.org/indicator/CM.MKT.LCAP.CD/} market capitals globally in 2020, attracts the attention of hundreds of millions of investors to pursue desirable financial assets and achieve investment goals. Recent years have witnessed significant development of quantitative trading \cite{an2022deep}, due to its instant and accurate order execution, and capability of analyzing and processing large amount of temporal financial data. Especially, intraday trading\footnote{https://www.investopedia.com/articles/trading/05/011705.asp}, where traders actively long/short pre-selected financial assets (at least a few times per day) to seize intraday trading opportunities, becomes one of the most profitable and risky quantitative trading tasks with the growth of discount brokerages (lower commission fee).

\begin{figure}[t]
    \centering
    \includegraphics[width=0.47\textwidth]{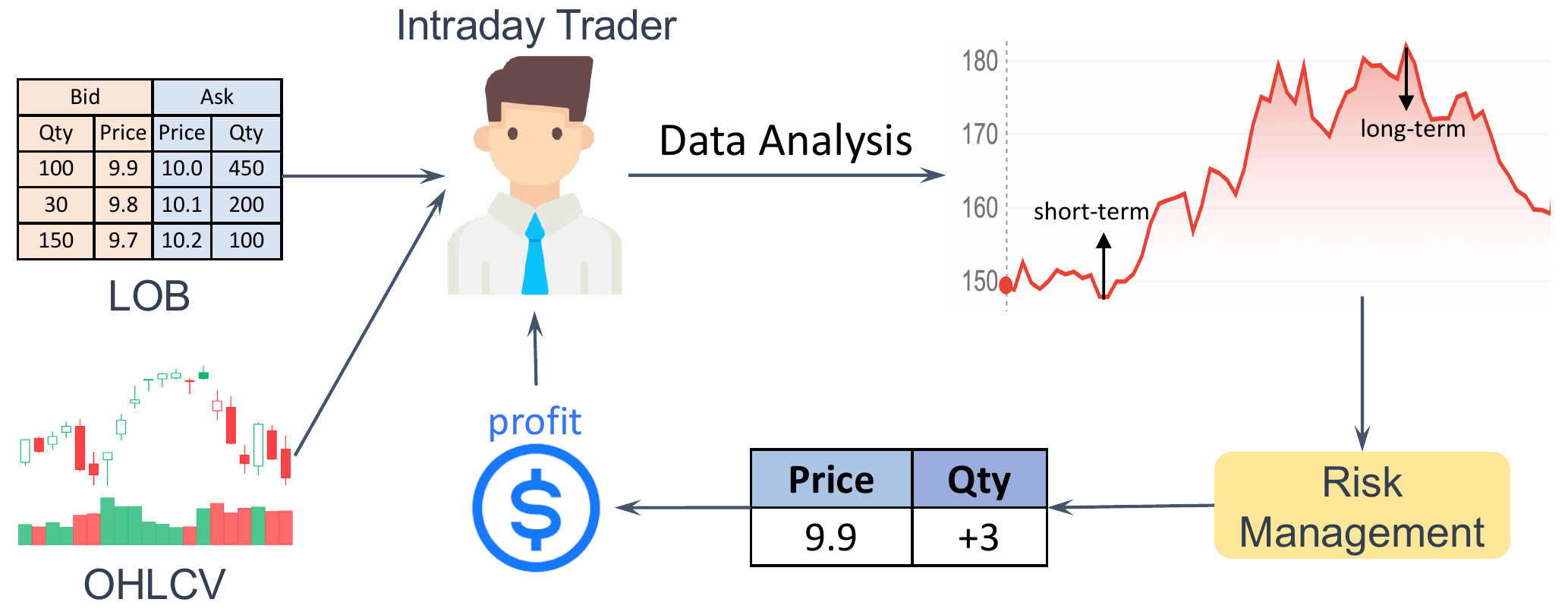}
    \caption{Workflow of professional human intraday trader}
    \label{fig:intraday_pipeline}
\end{figure}
Traditional intraday trading strategies use finance knowledge to discover trading opportunities with heuristic rules. For instance, momentum \citep{moskowitz2012time} trading is designed based on the assumption that the trend of financial assets in the past has the tendency to continue in the future. Mean reversion \citep{bollinger2002bollinger} focusing on the investment opportunities at the turning points. However, rule-based traditional methods exhibit poor generalization ability and only perform well in certain market conditions \citep{deng2016deep}. Another paradigm is to trade based on financial prediction. Many advanced machine learning models such as GCN \cite{feng2019temporal}, Transformer \cite{ding2020hierarchical} and LGBM \citep{ke2017lightgbm} have been introduced for predicting future prices \citep{ding2015deep}. Many other data sources such as economic news \citep{hu2018listening}, frequency of price \citep{zhang2017stock}, social media \citep{xu2018stock} and investment behaviors \citep{chen2019investment} have been added as additional information to further improve prediction performance. However, the high volatility and noisy nature of the financial market make it extremely difficult to accurately predict future prices \citep{fama1970efficient}. Furthermore, there is a noticeable gap between prediction signals and profitable trading actions \citep{feng2019temporal}. Thus, the overall performance of prediction-based methods is not satisfying as well.

Similar to other application scenarios of reinforcement learning (RL), quantitative trading also interacts with the environment (financial market) and maximizes the accumulative profit. Recently, RL has been considered as an attractive approach to quantitative trading as it allows training agents to directly output profitable trading actions with better generalization ability across various market conditions \cite{liu2020adaptive}.
Although there have been many successful RL-based quantitative trading methods \cite{ye2020reinforcement,wang2020commission,fang2021universal}, a vast majority of existing methods focus on the relatively low-frequency trading scenarios (e.g., day-level) and fail to capture the fleeting intraday investment opportunities. To design profitable RL methods for intraday trading, there are two major challenges. First, different from the low-frequency scenarios, intraday trading involves a much larger high-dimensional fine-grained action space to represent the price and quantity of orders for more accurate control of the financial market. Second, we need to learn meaningful multi-modality intraday market representation, which takes macro-level market, micro-level market and risk into consideration.


Considering the workflow of a professional human intraday trader (Figure \ref{fig:intraday_pipeline}), the trader first collects both micro-level and macro-level market information to analyze the market status. Then, he predicts the short-term and long-term price trend based on the market status. Later on, he conducts risk management and makes final trading decisions (when and how much to long/short at what price). Among many successful trading firms, this workflow plays a key role for designing robust and profitable intraday trading strategies. Inspired by it, we propose DeepScalper, a novel RL framework for intraday trading to tackle the above challenges by mimicking the information collection, short-term and long-term  market analysis and risk management procedures of human intraday traders. Our main contributions are three-fold:

\begin{itemize}
    \item We apply the dueling Q-Network with action branching to effectively optimize intraday trading agents with high-dimensional fine-grained action space. A novel reward function with hindsight bonus is designed to encourage RL agents making trading decisions with a long-term horizon of the entire trading day.
    \item We propose an multi-modality encoder-decoder architecture to learn meaningful temporal intraday market embedding, which incorporates both micro-level and macro-level market information. Furthermore, we design a risk-aware auxiliary task to keep balance between profit and risk.
    \item Through extensive experiments on real-world market data spanning over three years on six financial futures, we show that DeepScalper significantly outperforms many state-of-the-art baseline methods in terms of four financial criteria and demonstrate DeepScalper's practical applicability to intraday trading with a series of exploratory and ablative studies.
\end{itemize}

\section{Related Work}

\subsection{Traditional Finance Methods}
Technical analysis \citep{murphy1999technical}, which believes that past price and volume data have the ability to reflect future market conditions \citep{fama1970efficient}, is the foundation of traditional finance methods. Millions of technical indicators are designed to generate trading signals \citep{kakushadze2016101}. Momentum \citep{hong1999unified} and mean reversion \citep{poterba1988mean} are two well-known types of traditional finance methods based on technical analysis. Momentum trading assumes the trend of financial assets in the past has the tendency to continue in the future. Time Series Momentum \citep{moskowitz2012time} and Cross Sectional Momentum \citep{jegadeesh2002cross} are two classic momentum strategies. In contrast, mean reversion strategies such as Bollinger bands \citep{bollinger2002bollinger} assume that the price of financial assets will finally revert to the long-term mean. 

However, traditional finance methods are not perceptive enough to capture fleeting intraday patterns and only perform well in certain market conditions \citep{deng2016deep}. In recent years, many advanced machine learning methods have significantly outperformed traditional finance methods.

\subsection{Prediction-Based Methods}
As for prediction-based methods, they first formulate quantitative trading as a supervised learning task to predict the future return (regression) or price movement (classification). Later on, trading decisions are generated by the prediction results with a heuristic strategy generator (e.g., top-k in \cite{yoo2021accurate}). Specifically, \citet{wang19} combine attention mechanism with LSTM to model correlated time steps. To improve the robustness of LSTM, \citet{feng2018enhancing} apply adversarial training techniques for stock prediction. \citet{zhang2017stock} propose a novel State Frequency Memory (SFM) recurrent network with Discrete Fourier Transform (DFT) to discover multi-frequency patterns in stock markets. \citet{liu2020multi} introduce a multi-scale two-way neural network to predict the stock trend.
\citet{sun2022quantitative} propose an ensemble learning framework to train mixture of trading experts.


However, the high volatility and noisy nature of the financial market make it extremely difficult to accurately predict future prices \citep{fama1970efficient}. Furthermore, there is a noticeable gap between prediction signals and profitable trading actions \citep{feng2019temporal}. Thus, the overall performance of prediction-based methods is not satisfying as well.

\subsection{Reinforcement Learning Methods}
Recent years have witnessed the successful marriage of reinforcement learning and quantitative trading as RL allows training agents to directly output profitable trading actions with better generalization ability across various market conditions \cite{sun2021reinforcement}. \citet{neuneier1996optimal} make the first attempt to learn trading strategies using Q-learning. \citet{moody1999reinforcement} propose a policy-based method, namely recurrent reinforcement learning (RRL), for quantitative trading. However, traditional RL approaches have difficulties in selecting market features and learning good policy in large scale scenarios. To tackle these issues, many deep RL approaches have been proposed to learn market embedding through high dimensional data. \citet{jiang2017deep} use DDPG to dynamically optimize cryptocurrency portfolios. \citet{deng2016deep} apply fuzzy learning and deep learning to improve financial signal representation. \citet{yu2019model} propose a model-based RL framework for daily frequency portfolio trading. \citet{liu2020adaptive} propose an adaptive DDPG-based framework with imitation learning. \citet{ye2020reinforcement} proposed a State-Augmented RL (SARL) framework based on DPG with financial news as additional states. 

Although there are many efforts on utilizing RL for quantitative trading, a vast majority of existing RL methods focus on the relatively low-frequency scenarios (e.g., day-level) and fail to capture the fleeting intraday investment opportunities. We propose DeepScalper to fill this gap by mimicking the workflow of human intraday traders.

\section{Problem Formulation}
In this section, we first introduce necessary preliminaries and the objective of intraday trading. Next, we provide a Markov Decision Process (MDP) formulation of intraday trading.

\subsection{Intraday Trading}
Intraday trading is a fundamental quantitative trading task, where traders actively long/short \(\mathit{one}\) pre-selected financial asset within the same trading day to maximize future profit. Below are some necessary definitions for understanding the problem:

\theoremstyle{definition}
\begin{definition}{(OHLCV)}
\label{ohlcv}
OHLCV is a type of bar chart directly obtained from the financial market. OHLCV vector at time \(t\) is denoted as \(\textbf{x}_{t}=(p_{t}^{o}, p_{t}^{h}, p_{t}^{l}, p_{t}^{c}, v_{t})\), where \(p_{t}^{o}\) is the open price, \(p_{t}^{h}\) is the high price, \(p_{t}^{l}\) is the low price, \(p_{t}^{c}\) is the close price and \(v_{t}\) is the volume. 
\end{definition}


\theoremstyle{definition}
\begin{definition}{(Technical Indicator)}
\label{ti}
A technical indicator indicates a feature calculated by a formulaic combination of the original OHLCV to uncover the underlying pattern of the financial market. We denote the technical indicator vector at time \(t\): \(\textbf{y}_t = {\textstyle \bigcup_{k}y_t^{k}} \), where \(y_{t}^{k} = f_k(\textbf{x}_{t-h},...,\textbf{x}_{t},\theta^k )\), \(\theta^k\) is the parameter of technical indicator \(k\).
\end{definition}

\theoremstyle{definition}
\begin{definition}{(Limit Order)}
\label{lo}
A limit order is an order placed to long/short a certain number of shares at a specific price. It is defined as a tuple \(l=(p_{target},\pm q_{target})\), where \(p_{target}\) represents the submitted target price, \(q_{target}\) represents the submitted target quantity, and \(\pm\) represents the trading direction (long/short).
\end{definition}

\begin{figure}[t]
    \centering
    \includegraphics[width=0.47\textwidth]{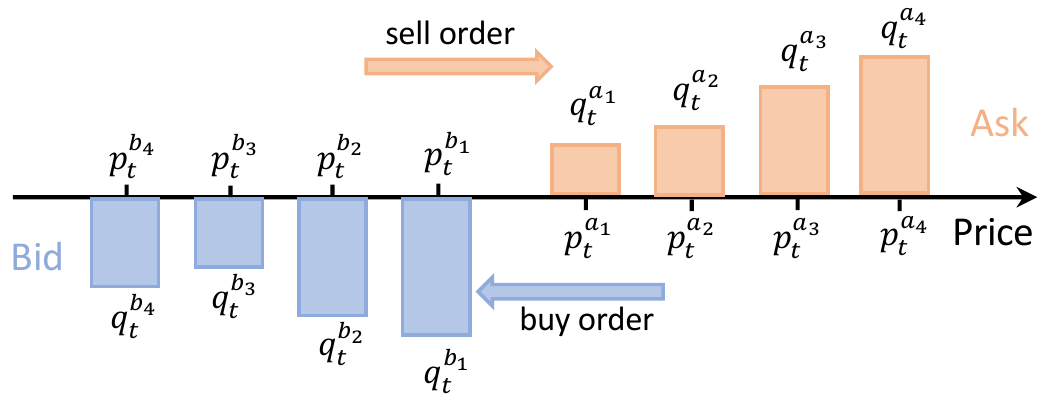}
    \caption{A snapshot of 4-level limit order book (LOB)}
    \label{fig:lob}
\end{figure}

\theoremstyle{definition}
\begin{definition}{(Limit Order Book)}
\label{lob}
As shown in Figure \ref{fig:lob}, a limit order book (LOB) contains public available aggregate information of limit orders by all market participants. It is widely used as market microstructure \citep{madhavan2000market} in finance to represent the relative strength between buy and sell side. We denote an \(m\) level LOB at time \(t\) as \(\textbf{b}_{t}=(p_{t}^{b_1},p_{t}^{a_1},q_{t}^{b_1},q_{t}^{a_1},...,p_{t}^{b_m},p_{t}^{a_m},q_{t}^{b_m},q_{t}^{a_m})\), where \(p_{t}^{b_i}\) is the level \(i\) bid price, \(p_{t}^{a_i}\) is the level \(i\) ask price, \(q_{t}^{b_i}\) and \(q_{t}^{a_i}\) are the corresponding quantities.
\end{definition}

\theoremstyle{definition}
\begin{definition}{(Matching System)}
\label{ms}
The matching system is an electronic system that matches the buy and sell orders for the financial market. It is usually used to execute orders for different traders in the exchange.
\end{definition}

\theoremstyle{definition}
\begin{definition}{(Position)}
\label{position}
Position \({pos}_t\) at time \(t\) is the amount and direction of a financial asset owned by traders. It represents a long (short) position when \({pos}_t\) is positive (negative).
\end{definition}

\theoremstyle{definition}
\begin{definition}{(Net Value)}
\label{nv}
Net value is the normalised sum of cash and value of financial assets held by a trader. The net value at time \(t\) is denoted as \(n_t = (c_t + p^c_t\times\left|{pos}_t\right|)/c_1\), where \(c_t\) is the cash at time \(t\) and \(c_1\) is the initial amount of cash.
\end{definition}
\noindent
In real-world intraday trading, traders are allocated some cash into the account at the beginning of each trading day. During the trading time, traders analyze the market and make their trading decisions. Then, they submit their orders (desired price and quantity) to the matching system. The matching system will execute orders at best available price (possibly at multiple price when market liquidation is not enough for large orders) and then return execution results to traders. At the end of the trading day, traders close all remaining positions at market price to avoid overnight risk and hold 100\% cash again. The objective of intraday trading is to maximize accumulative profit for a period of multiple continuous trading days (e.g., half a year).

Comparing to conventional low-frequency trading scenarios, intraday trading is more challenging since intraday traders need to capture the \textit{tiny} price fluctuation with much \textit{less} responsive time (e.g., 1 min). In addition, intraday trading involves a large fine-grained trading action space that represents a limit order to pursue more accurate control of the market. 

\subsection{MDP Formulation}

We formulate intraday trading as a MDP, which is constructed by a 5-tuple \((\mathcal{S},\mathcal{A} ,T, R,\gamma)\). Specifically, \(\mathcal{S}\) is a finite set of states. \(\mathcal{A}\) is a finite set of actions. \(T:\mathcal{S}\times\mathcal{A} \times \mathcal{S}\longrightarrow [0,1]\) is a state transaction function, which consists of a set of conditional transition probabilities between states. \(R:\mathcal{S}\times\mathcal{A} \longrightarrow \mathcal{R}\) is the reward function, where \(\mathcal{R}\) is a continuous set of possible rewards and \(R\) indicates the immediate reward of taking an action in a state. \(\gamma\in[0,1)\) is the discount factor. A (stationary) policy $\pi: \mathcal{S} \times \mathcal{A}\longrightarrow [0,1]$ assigns each state $s \in \mathcal{S}$ a distribution over actions, where $a \in \mathcal{A}$ has probability $\pi(a|s)$. In intraday trading, \(\mathcal{O}, \mathcal{A}, R\) are set as follows.


\textbf{State.}
Due to the particularity of the financial market, the state \(s_t \in \mathcal{S}\) at time \(t\) can be divided into three parts: macro-level market state \(s^a_t \in \mathcal{S}^a\), micro-level market state \(s^i_t \in \mathcal{S}^i\) and trader's private state set  \(s^p_t \in \mathcal{S}^p\). Specifically, we use a vector \(\textbf{y}_\textbf{t}\) of 11 technical indicators and the original OHLCV vector \(\textbf{x}_\textbf{t}\) as macro-level state following \cite{yoo2021accurate}, the historical LOB sequence \((\textbf{b}_\textbf{t-h},...,\textbf{b}_\textbf{t})\) with length \(h+1\) as micro-level state and trader's private state \(\textbf{z}_\textbf{t}=({pos}_t, c_t, t_t)\), where \({pos}_t\), \(c_t\) and \(t_t\) are the current position, cash and remaining time. The entire set of states can be denoted as \(\mathcal{S}=(\mathcal{S}^a, \mathcal{S}^i, \mathcal{S}^p)\). Compared to previous formulations, we introduce the LOB and trader's private state as additional information to effectively capture intraday trading opportunities. 

\begin{figure*}[ht]
\begin{center}
\includegraphics[width=0.99\textwidth]{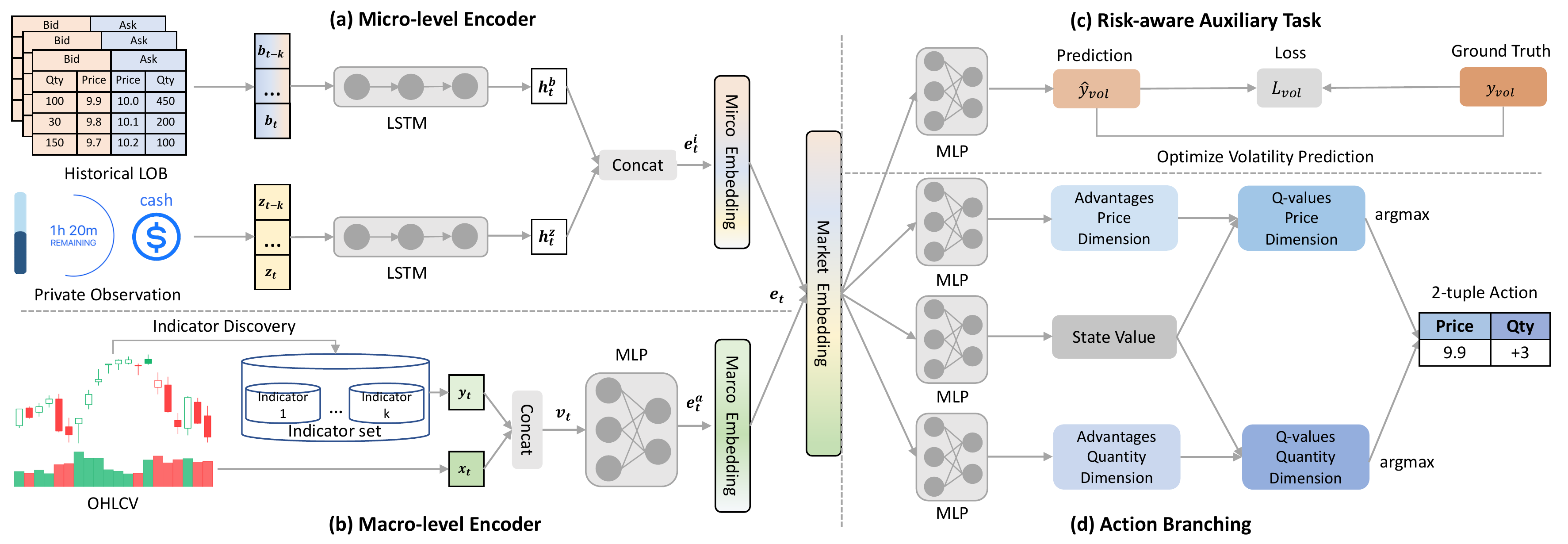}
\caption{An overview of the proposed DeepScalper framework. We show four individual building blocks: (a) micro-level encoder, (b) macro-level encoder, (c) risk-aware auxiliary task, and (d) RL optimization with action branching.}
\label{fig:deepscalper}
\end{center}
\end{figure*}
\textbf{Action.}
Previous works \citep{deng2016deep,liu2020adaptive} lie in low-frequency trading scenarios, which generally stipulate that the agent trades a fixed quantity at market price and applies a coarse action space with three options (long, hold, and short). However, when focusing on relatively high-frequency trading scenarios (e.g., intraday trading), tiny price fluctuation (e.g., 1 cent) is of vital importance to final profit that makes the market price execution and fixed quantity assumptions unacceptable. In the real-world financial market, traders have the freedom to decide both the target price and the quantity by submitting limit orders. We use a more realistic two-dimensional fine-grained action space for intraday trading, which represents a limit order as a tuple \((p_{target},\pm q_{target})\). 
\(p_{target}\) is the target price, \(q_{target}\) is the target quantity and \(\pm\) is the trading direction (long/short). It is also worth pointing out that when the quantity is zero, it indicates that we skip the current time step with no order placement.

\textbf{Reward.}
We define the reward function as the change of account P\&L, which shows the value fluctuation (profit \& loss) of the account:
\[r_t = \underbrace{(p^c_{t+1} - p^c_{t}) \times {pos}_{t}}_{instant \ profit} - \underbrace{\delta \times p^c_{t}\times\left | {pos}_{t}-{pos}_{t-1} \right | }_{transaction \ fee}\]
where \(p^c_{t}\) is the close price at time \(t\), \(\delta\) is the transaction fee rate and \({pos}_t\) is the position at time \(t\).





\section{DeepScalper}

In this section, we introduce DeepScalper (an overview in Figure \ref{fig:deepscalper}) for intraday trading. We describe the four components of DeepScalper: 1) RL optimization with action branching; 2) reward function with a hindsight bonus; 3) intraday market embedding; 4) risk-aware auxiliary task orderly.

\subsection{RL Optimization with Action Branching}
Comparing to conventional low-frequency trading scenarios, intraday trading tries to seize the fleeting \textit{tiny} price fluctuation with much \textit{less} responsive time. To provide more accurate trading decisions, intraday trading involves a much larger two-dimensional (price and quantity) fine-grained action space. However, learning from scratch for tasks with large action spaces remains a critical challenge for RL algorithms \cite{zahavy2018learn,bellemare2017distributional}. For intraday trading, while human traders can usually detect the subset of feasible trading actions in a given market condition, RL agents may attempt inferior actions, thus wasting computation time and leading to capital loss.    

As possible intraday trading actions can be naturally divided into two components (e.g., desired price and quantity), we propose to adopt the Branching Dueling Q-Network (BDQ) \citep{tavakoli2018action} for decision-making. Particularly, as shown in Figure \ref{fig:deepscalper}(d), 
BDQ distributes the representation of the state-dependent action advantages in both the price and quantity branches. Later, it simultaneously adds a single additional branch to estimate the state-value function. Finally, the advantages and the state value are combined via an aggregating layer to output the Q-values for each action dimension. During the inference period, these Q-values are then queried with argmax to generate a joint action tuple to determine the final trading actions. 

Formally, intraday trading is formulated as a sequential decision making problem with two action dimensions of \(|p|=n_{p}\) discrete relative price levels and \(|q|=n_{q}\) discrete quantity proportions. The individual branch's Q-value \(Q_{d}\) at state \(s \in S\) and the action \(a_{d} \in \mathcal{A}_{d}\) are expressed in terms of the common state value \( V(s) \) and the corresponding (state-dependent) action advantage \citep{wang2016dueling} \( Adv_{d}(s,a_{d}) \) for \(d \in \{p,q\}\):

\[ Q_{d}(s,a_{d})=V(s) + (Adv_{d}(s,a_{d})-\frac{1}{n}\sum_{a_{d}^{'} \in A_{d}}^{}Adv_{d}(s,a_{d}^{'}) ) \]

We train our Q-value function approximator as Q-Network with parameter \(\theta_q\) based on the one-step temporal-difference learning with target $y_d$ in a recursive fashion:

\[ y_{d} = r + \gamma \max_{a_{d}'\in A_{d}}Q_{d}^{-}(s',a_{d}',\theta_{q}) ),d\in\{p,q\} \]
where \(Q_{d}^-\) denoting the branch \(d\) of the target network \(Q^-\), \(r\) is the reward function result and \(\gamma\) is the discount factor.

Finally, we calculate the following loss function:
\[ L_{q}(\theta_q) =\mathit{E}_{(s,a,r,s')\sim D}[\frac{1}{N}\sum_{d\in\{p,q\}}(y_{d}-Q_{d}(s,a_{d},\theta_q))^{2}] \]

where \(D\) denotes a prioritized experience replay buffer. \(a\) denotes the joint-action tuple \((p,q)\). By differentiating the Q-Network loss function with respect to \(\theta_q\), we get the following gradient: 
\begin{gather*}
\bigtriangledown_{\theta_q}L_q(\theta_q)=\mathit{E}_{(s,a,r,s^{'})\sim D}[(r + \gamma \max_{a_{d}^{'}\in A_{d}}Q_{d}(s^{'},a_{d}^{'},\theta_{q})\\
-Q_{d}(s,a_{d},\theta_q))\bigtriangledown_{\theta_q}Q_{d}(s,a_{d},\theta_q)]
\end{gather*}

In practice, we optimize the loss function by stochastic gradient descent, rather than computing the full expectations in the above gradient, to maintain computational efficiency.




\begin{figure}[th]
\begin{center}
\includegraphics[width=0.99\columnwidth]{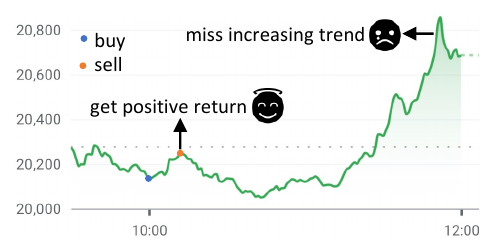}
\caption{Illustration of the motivation of the hindsight bonus}
\label{fig:hindsight_idea}
\end{center}
\end{figure}

\subsection{Reward Function with Hindsight Bonus}

One major issue for training directly with the profit \& loss reward is that RL agents tend to pay too much attention to the short-term price fluctuation \citep{wang2021deeptrader}. Although the agent performs well in capturing local trading opportunities, ignoring the overall long-term price trend could lead to significant loss. Here, we design a novel reward function with a hindsight bonus to tackle this issue. 
To demonstrate the motivation of the hindsight bonus, considering a pair of buy/sell actions in  Figure \ref{fig:hindsight_idea}, the trader feels happy at the point of selling the stock, since the price of the stock increases. However, this sell decision is actually a bad decision in the long run. The trader feels disappointed before 12:00 since he/she misses the main increasing wave due to the short horizon. It is more reasonable for RL agents to evaluate one trading action from both short-term and long-term perspectives. Inspired by this, we add a hindsight bonus, which is the expected profit for holding the assets for a longer period of \(h\) with a weight term \(w\), into the reward function to add a long-term horizon while training intraday RL agents:  

\[r_t^{hind} = r_t + \underbrace{w \times (p^c_{t+h} - p^c_{t})\times {pos}_t}_{hindsight \ bonus}\]
\noindent
where \(p^c_{t}\) is the close price at time \(t\), \(w\) is the weight of the hindsight bonus, \(h\) is the horizon of the hindsight bonus and \({pos}_t\) is the position at time \(t\). 

Noticeably, we only use the reward function with a hindsight bonus for training to better understand the market. During the test period, we continue to use the original reward \(r_t\) to calculate the profits. Furthermore, the hindsight reward function somehow ignores details of price fluctuation between \(t+2\) to \(t+h-1\) and focuses on the trend of this period, which is computational efficient and shows robust performance in practice.

\subsection{Intraday Market Embedding}
To learn a meaningful multi-modality intraday market embedding, we propose an encoder-decoder architecture to represent the market from the micro-level and macro-level, respectively. 

For micro-level encoder, we choose LOB data and trader's private state to learn the micro-level market embedding. LOB is widely used to analyze the relative strength of the buy and sell side based on micro-level trading behaviors, and private state of traders is considered insightful to capture micro-level trading opportunities \citep{nevmyvaka2006reinforcement}. At time \(t\), we have a sequence of historical LOB embeddings \((\textbf{b}_\textbf{t-k},...,\textbf{b}_\textbf{t})\) and trader's private state embedding \((\textbf{z}_\textbf{t-k},...,\textbf{z}_\textbf{t})\), where \(k+1\) is the sequence length. As shown in Figure \ref{fig:deepscalper}(a), we feed them into two different LSTM layers and concatenate the last hidden states \(\textbf{h}_{\textbf{t}}^{\textbf{b}}\) and \(\textbf{h}_{\textbf{t}}^{\textbf{z}}\) of the two LSTM layers as the micro-level embedding \(\textbf{e}^\textbf{i}_\textbf{t}\) at time \(t\).

For macro-level encoder, we pick raw OHLCV data and technical indicators to learn the macro-level embedding. The intuition here is that OHLCV reflects the original market status, and technical indicators offer additional information. At time \(t\), we firstly concatenate OHLCV vector \(\textbf{x}_\textbf{t}\) and technical indicator vector \(\textbf{y}_\textbf{t}\) as input \(\textbf{v}_\textbf{t}\). As shown in Figure \ref{fig:deepscalper}(b), the concatenated embedding is then fed into a multilayer perceptron (MLP). The MLP output is applied as the macro-level embedding \(\textbf{e}^\textbf{a}_\textbf{t}\) at time \(t\). Finally, we concatenate micro-level embedding and macro-level embedding together as the market embedding \( \textbf{e}_\textbf{t} \). Our market embedding is better than that of previous work, since it incorporates the micro-level market information. 

\subsection{Risk-Aware Auxiliary Task}
As risk management is of vital importance for intraday trading, we propose a risk-aware auxiliary task by predicting volatility to take into account market risk as shown in Figure \ref{fig:deepscalper}(c). Volatility is widely used as a coherent measure of risk that describing the statistical dispersion of returns in finance \citep{bakshi2003delta}. We analyze the reasons why volatility prediction is an effective auxiliary task to improve the trading policy learning as follows.

First, it is consistent with the general trading goal, which is to maximize long-term profit under certain risk tolerance. Second, future volatility is easier to predict compared to future price. For instance, considering the day that the president election result will be announced, nobody can know the result in advance, which will lead the stock market to increase or decrease. However, everyone knows that there would be a huge fluctuation in the stock market, which increases future volatility. Third, predicting future price and volatility are two closely related tasks. Learning value function approximation and volatility prediction simultaneously can help the agent learn a more robust market embedding. The definition of volatility is the variance of return sequence \(y_{vol} = \sigma(\textbf{r}) \),
where \(\textbf{r}\) is the vector of return at each time step. Volatility prediction is a regression task with market embedding \(\textbf{e}_\textbf{t}\) as input and \(y_{vol}\) as target. We feed the market embedding into a single layer MLP with parameters \(\theta_v\). The output \(\hat{y_{vol}}\) is the predicted volatility. We train the network by minimizing the mean squared error.
\[\hat{y_{vol}} =MLP(e_t,\theta_v) \] 
\[L_{vol}(\theta_v)=(y_{vol}-\hat{y_{vol}})^2\]
The overall loss function is defined as:
\[L=L_q + \eta*L_{vol}\]
where $L_q$ is the Q value loss and \(\eta\) is the relative importance of the auxiliary task.

\section{Experiment Setup}
\begin{table}[t]
\setlength\tabcolsep{4pt}
    \centering
    \begin{tabular}{|c|ccccc|}
    \hline
    \textbf{Dataset}  & \textbf{Freq} & \textbf{Number} & \textbf{Days} & \textbf{From} & \textbf{To} \\ \hline
    Stock index  & 1min & 2 & 251 & 19/05/01 & 20/04/30 \\
    Treasury bond & 1min & 4 & 662 & 17/11/29 & 20/07/17 \\
    \hline
    \end{tabular}
    \caption{Dataset statistics detailing data frequency, number of financial assets, trading days and chronological period}
    \label{tab:dataset}
\end{table}
\subsection{Datasets and Features}
To conduct a comprehensive evaluation of DeepScalper, we evaluate it on \(\mathit{six}\) financial assets from \(\mathit{two}\) real-world datasets (stock index and treasury bond) spanning over \(\mathit{three}\) years in the Chinese market collected from Wind\footnote{https://www.wind.com.cn/}. We summarize the statistics of the two datasets in Table \ref{tab:dataset} and further elaborate on them as follows:

\(\textbf{Stock index}\) is a dataset containing the minute-level OHLCV and 5-level LOB data of two representative stock index futures (IC and IF) in the Chinese market. IC is a stock index future calculated based on 500 small and medium market capitalization stocks. IF is another stock index future that focuses on the top 300 large capitalization stocks. For each stock index future, we split the dataset with May-Dec, 2019 for training and Jan-April, 2020 for testing. 

\(\textbf{Treasury bond}\) is a dataset containing the minute-level OHLCV and 5-level LOB data of four treasury bond futures (T01, T02, TF01, TF02). These treasury bond futures are mainstream treasury bond futures with the highest liquidity in the Chinese market. For each treasury bond, we use 2017/11/29 - 2020/4/29 for training and 2020/04/30 - 2020/07/17 for testing.

To describe macro-level financial markets, we generate 11 temporal features from the original OHLCV as shown in Table \ref{tab:feature} following \cite{yoo2021accurate}. $z_{open}$, $z_{high}$ and $z_{low}$ represent the relative values of the open, high, and low prices compared to the close price at the current time step, respectively. $z_{close}$ and $z_{adj\_close}$ represent the relative values of the closing and adjusted closing prices compared to the time step \(t-1\). $z_{dk}$ represents a long-term moving average of the adjusted close prices during the last \(k\) time steps compared to the current close price. For micro-level markets, we extract a 20-dimensional feature vector from the 5-level LOB where each level contains bid, ask price and bid, ask quantity following \cite{wang2020commission}.

\begin{table}[t]
\setlength\tabcolsep{11pt}
    \newcommand{\tabincell}[2]{\begin{tabular}{@{}#1@{}}#2\end{tabular}}
    \centering
    \begin{tabular}{|l|l|}
    \hline
    \textbf{Features}  & \textbf{Calculation Formula}  \\ \hline
    $z_{open},z_{high},z_{low}$ & e.g., $ z_{open} = open_t/close_t - 1$ \\ 
    $z_{close}, z_{adj\_close}$ & e.g., $ z_{close} = close_t / close_{t-1} - 1$\\ \hline
    \tabincell{l}{$z_{d\_5}, z_{d\_10}, z_{d\_15}$ \\$z_{d\_20}, z_{d\_25}, z_{d\_30}$} & e.g., \(z_{d\_5} = \frac{{\textstyle \sum_{i=0}^{4}adj\_close_{t-i}/5}}{\textstyle adj\_close_{t}}-1\)  \\ \hline
    \end{tabular}
    \caption{Features to describe macro-level financial markets}
    \label{tab:feature}
\end{table}

\subsection{Evaluation Metrics}
We evaluate DeepScalper on four different financial metrics, including one profit criterion and three risk-adjusted profit criteria:
\begin{itemize}
    \item \textbf{Total Return (TR)} is the overall return rate for the entire trading period.
    It is defined as \( TR = \frac{n_{t}-n_{1}}{n_{1}} \), where \({n}_{t}\) is the final net value and \({n}_{1}\) is the initial net value.
    \item \textbf{Sharpe Ratio (SR)} \cite{sharpe1994sharpe} considers the amount of extra return that a trader receives per unit of increased risk. It is defined as: \( SR = \mathbb{E}[\mathbf{r} ]/\sigma[\mathbf{r}] \), where \(\mathbf{r}\)  denotes the historical sequence of the return rate.
    \item \textbf{Calmar Ratio (CR)} is defined as \(CR = \frac{\mathbb{E}[\mathbf{r}]}{MDD}\). It is calculated as the expected return divided by the maximum drawdown (MDD) of the entire trading period, where MDD measures the largest loss from any peak to show the worst case.
    \item \textbf{Sortino Ratio (SoR)} applies the downside deviation (DD) as the risk measure. It is defined as: \(SoR = \frac{\mathbb{E}[\mathbf{r}]}{DD}\), where DD is the variance of the negative return. 
\end{itemize}

\subsection{Baseline}
We compare DeepScalper with nine baseline methods consisting of three traditional finance methods, three prediction-based methods, and three reinforcement learning methods.

\noindent
\textit{Traditional Finance Methods}
\begin{itemize}
    \item \textbf{Buy \& Hold (BAH)}, which is usually used to reflect the market average, indicates the trading strategy that buys the pre-selected financial assets with full position at the beginning and holds until the end of the trading period.
    \item \textbf{Mean Reversion (MV)} \citep{poterba1988mean} is a traditional finance method designed under the assumption that the price of financial assets will eventually revert to the long-term mean. In practice, it shows stellar performance under volatile market conditions.  
    \item \textbf{Time Series Momentum (TSM)} \citep{moskowitz2012time} is an influential momentum-based method, which long (short) financial assets with increasing (decreasing) trend in the past. This is in line with the principle that the stronger is always stronger in the financial market. 
\end{itemize}

\noindent
\textit{Prediction Based Methods}
\begin{itemize}
    \item \textbf{MLP} \citep{schierholt1996stock} use the classic multi-layer perceptron for future return prediction. We apply a three-layer MLP with hidden size 128.
    \item \textbf{GRU} \citep{chung2014empirical} use a newer generation of recurrent networks with gated recurrent units for future return prediction. We apply a two-layer GRU with hidden size 64. 
    \item \textbf{LGBM} \citep{ke2017lightgbm} is an efficient implementation of the gradient boosting decision tree with gradient-based one-side sampling and exclusive feature bundling.
\end{itemize}

\noindent
\textit{Reinforcement Learning Methods}
\begin{itemize}
    \item \textbf{DQN} \citep{zhang2020deep} applies the deep Q-network with a novel state representation and reward function for quantitative trading, which shows stellar performance in more than 50 financial assets.
    \item \textbf{DS-NH} is a variant of DeepScalper (DS), which removes the hindsight bonus from the reward function. 
    \item \textbf{DS-NA} is a variant of DeepScalper (DS), which removes the risk-aware auxiliary task.
\end{itemize}

\subsection{Preprocessing and Experiment Setup}
For macro-level features, we directly calculate the 11 technical indicators following the formulas in Table \ref{tab:feature}. For micro-level features, we divide the order price and quantity of each level by the first-level price and quantity, respectively, for normalization. For missing values, we fill the empty price with the previous one and empty quantity as zero to maintain the consistency of time series data. To make the evaluation more realistic, we further consider many practical real-world constraints. The transaction fee rate \(\delta\) is set as \(2.3\times10^{-5}\) and \(3\times10^{-6}\) for stock index futures and treasury bond futures respectively, which is consistent with the real-world scenario\footnote{China Financial Futures Exchange: http://www.cffex.com.cn/en\_new/index.html}. Since leverage such as margin loans is widely used for intraday trading, we apply a fixed five-times leverage to amplify profit and volatility. Time is discretized into 1 min interval and we assume that the agent can only long/short a financial future at the end of each minute. The account of RL agents is initialized with enough cash to buy 50 shares of the asset at the beginning. The maximum holding position is 50.

We perform all experiments on a Tesla V100 GPU. Grid search is applied to find the optimal hyperprarameters. We explore the look-ahead horizon \(h\) in [30, 60, 90, 120, 150, 180], importance of hindsight bonus \(w\) in \([1e^{-3}, 5e^{-3}, 1e^{-2}, 5e^{-2}, 1e^{-1}]\) and importance of auxiliary task \(\eta\) in [0.5, 1.0]. As for neural network architectures, we search the hidden units of MLP layers and GRU layer in [32, 64, 128] with ReLU as the activation function. we use Adam as the optimizer with learning rate \(\alpha \in (1e^{-5},1e^{-3})\) and train DeepScalper for 5 epochs in all financial assets. Following the iterative training scheme in \citep{nevmyvaka2006reinforcement}, we augment traders' private state repeatedly during the training to improve data efficiency. We run experiments with 5 different random seeds and report the average performance. It takes 1.5 and 3.5 hours to train and test Deepscalper on each financial asset in the stock index and treasury bond datasets, respectively. As for other baselines, we use the default settings in their public implementations\footnote{Qlib: https://github.com/microsoft/qlib} \footnote{FinRL: https://github.com/AI4Finance-Foundation/FinRL}.

\begin{table*}[!thb]
\small
\centering

\resizebox{2.1\columnwidth}{!}{
\begin{tabular}{|c|c|cccc|cccc|}

\hline
\multicolumn{2}{|}{} & \multicolumn{4}{| c | }{Stock Index} & \multicolumn{4}{ c| }{Treasury Bond}  \\ 
\cline{1-10}
Type & Models & TR(\%)$\uparrow$ & SR$\uparrow$ & CR$\uparrow$ & SoR$\uparrow$ & TR(\%)$\uparrow$ &  SR$\uparrow$ &  CR$\uparrow$ & SoR$\uparrow$ \\
\hline
& BAH &  5.65 & 0.15 & 0.02 & 0.27 & -14.26 & -3.42 & -0.25 & -4.40 \\
FIN & MV   & 8.39 & 1.18 & 0.21 & 2.22 & -0.29 & -0.59 & -0.04 & -0.62 \\
& TSM  & -27.62 & -2.83 & -0.21 & -3.08 & -3.02 & -5.35 & -0.31 & -6.20 \\
\hline
& MLP & -0.73 $\pm$ 7.11 & -0.14 $\pm$ 1.02 & -0.01 $\pm$ 0.58 & -0.24 $\pm$ 1.77 & 0.59 $\pm$ 1.11 & 1.42 $\pm$ 1.30 & 0.42 $\pm$ 0.51 & 2.33 $\pm$ 1.42 \\ 
PRE & GRU  & 5.66 $\pm$ 4.98 & 1.25 $\pm$ 0.66 & 0.24 $\pm$ 0.18 & \textcolor[RGB]{255,0,185}{2.40 $\pm$ 0.82} & 1.02 $\pm$ 2.10 & 1.90 $\pm$ 1.72 & 0.55 $\pm$ 0.57 & 3.69 $\pm$ 2.09 \\
& LGBM & 7.62 $\pm$ 1.14 & 1.26 $\pm$ 0.22 & \textcolor[RGB]{255,0,185}{0.28 $\pm$ 0.05} & 1.59 $\pm$ 0.22 & 1.45 $\pm$ 0.17 & 2.43 $\pm$ 0.43 & 0.58 $\pm$ 0.09 & 3.68 $\pm$ 0.52 \\ \hline
& DQN & 7.74 $\pm$ 3.52 & 1.25 $\pm$ 0.62  & \textcolor[RGB]{255,0,185}{0.28 $\pm$ 0.17} & 1.79 $\pm$ 0.91 & 3.51 $\pm$ 1.05 & 4.01 $\pm$ 1.27 & 1.15 $\pm$ 0.39 & 5.66 $\pm$ 1.33 \\
RL & DS-NH & 8.17 $\pm$ 5.07 & 0.98 $\pm$ 0.77 & 0.17 $\pm$ 0.17 & 1.37 $\pm$ 0.88 & 3.38 $\pm$ 1.28 & \textcolor[RGB]{255,0,185}{4.42 $\pm$ 1.21} & \textcolor[RGB]{255,0,185}{1.39 $\pm$ 0.45} & 6.85 $\pm$ 1.19 \\
& DS-NA & \textcolor[RGB]{255,0,185}{9.74 $\pm$ 5.12} & \textcolor[RGB]{255,0,185}{1.32 $\pm$ 0.76} & 0.26 $\pm$ 0.21 & 2.19 $\pm$ 1.11 & \textcolor[RGB]{255,0,185}{4.17 $\pm$ 1.44} & 4.27 $\pm$ 0.99 & 1.38 $\pm$ 0.43 & \textcolor[RGB]{129,0,121}{7.59 $\pm$ 1.49} \\
& DS & \textcolor[RGB]{129,0,121}{$12.74^{\clubsuit} \pm 4.65$} & \textcolor[RGB]{129,0,121}{$1.76^{\clubsuit} \pm 0.61$} & \textcolor[RGB]{129,0,121}{$0.34^{\clubsuit} \pm 0.16$} & \textcolor[RGB]{129,0,121}{$2.58^{\clubsuit} \pm 0.72$} & \textcolor[RGB]{129,0,121}{$4.79^{\clubsuit} \pm 0.99$} & \textcolor[RGB]{129,0,121}{$4.75^{\clubsuit} \pm 1.25$} & \textcolor[RGB]{129,0,121}{$1.82^{\clubsuit} \pm 0.41$} & \textcolor[RGB]{255,0,185}{7.4 $\pm$ 1.22} \\ \hline
\multicolumn{2}{ |c | }{\% Improvement} & \textcolor[RGB]{0,162,96}{30.80$\uparrow$} & \textcolor[RGB]{0,162,96}{33.33$\uparrow$} & \textcolor[RGB]{0,162,96}{21.42$\uparrow$} & \textcolor[RGB]{0,162,96}{7.50$\uparrow$} & \textcolor[RGB]{0,162,96}{14.87$\uparrow$} & \textcolor[RGB]{0,162,96}{7.47$\uparrow$} & \textcolor[RGB]{0,162,96}{30.94$\uparrow$} & \textcolor[RGB]{215,60,60}{2.57$\downarrow$} \\ \hline
\end{tabular}
}

\captionof{table}{Profitability comparison (mean and standard deviation of 5 individual runs) with 9 baselines including traditional finance (FIN), prediction based (PRE) and reinforcement learning (RL) methods. All three FIN models are \textit{deterministic} methods without the performance standard deviation. Purple and pink show best \& second best results. $\clubsuit$ indicates improvement over SOTA baseline is statistically significant ($p<0.01$) under Wilcoxon's signed rank test.}
\label{fig:performance}
\end{table*}

\begin{figure}[thb]
  \centering
  \subfloat[{IC}\label{fig:tfm02}]{%
      \includegraphics[width=0.23\textwidth]{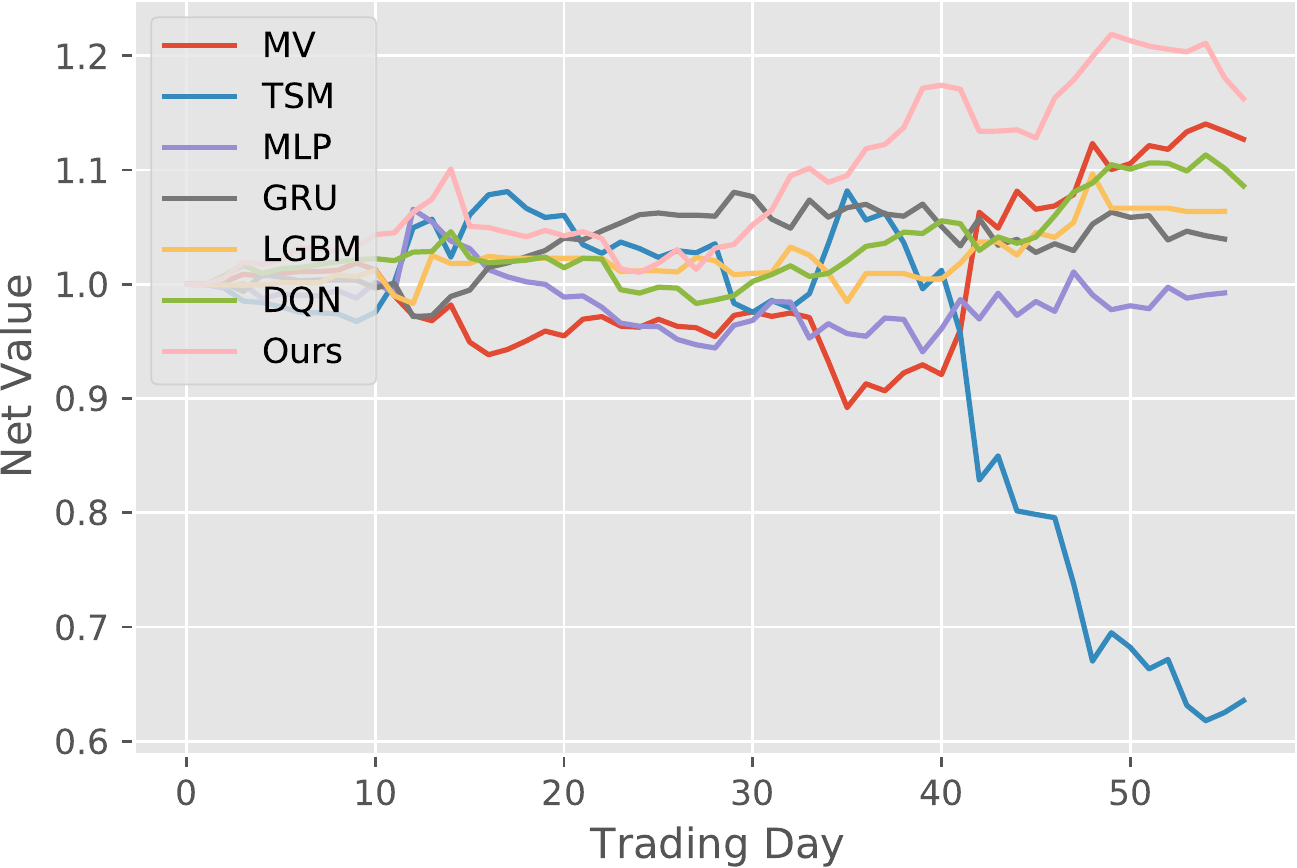}
     }
     \hfill
     \subfloat[{IF}\label{fig:tm02}]{%
      \includegraphics[width=0.23\textwidth]{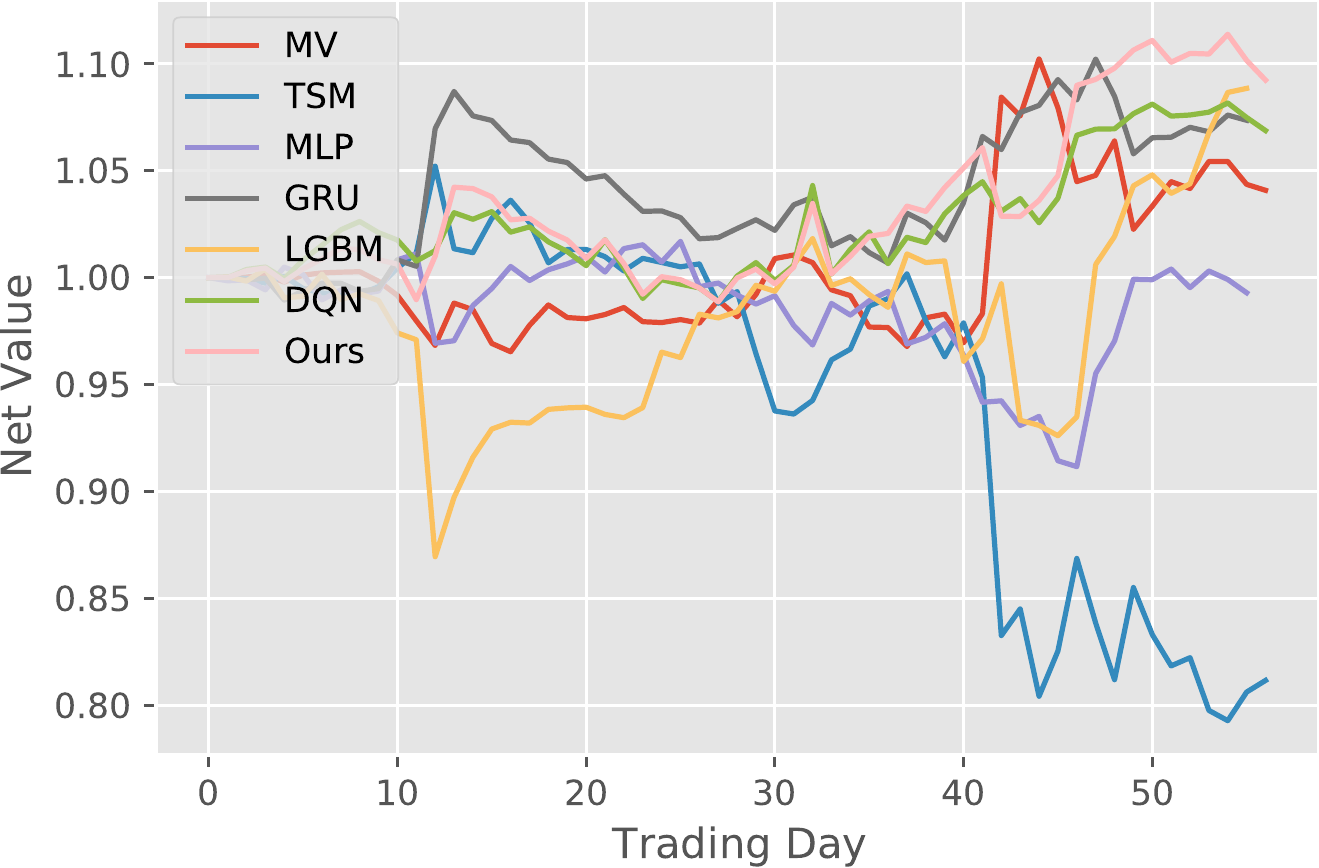}
     }
     \hfill
     \subfloat[{TF01}\label{fig:tfm02}]{%
      \includegraphics[width=0.23\textwidth]{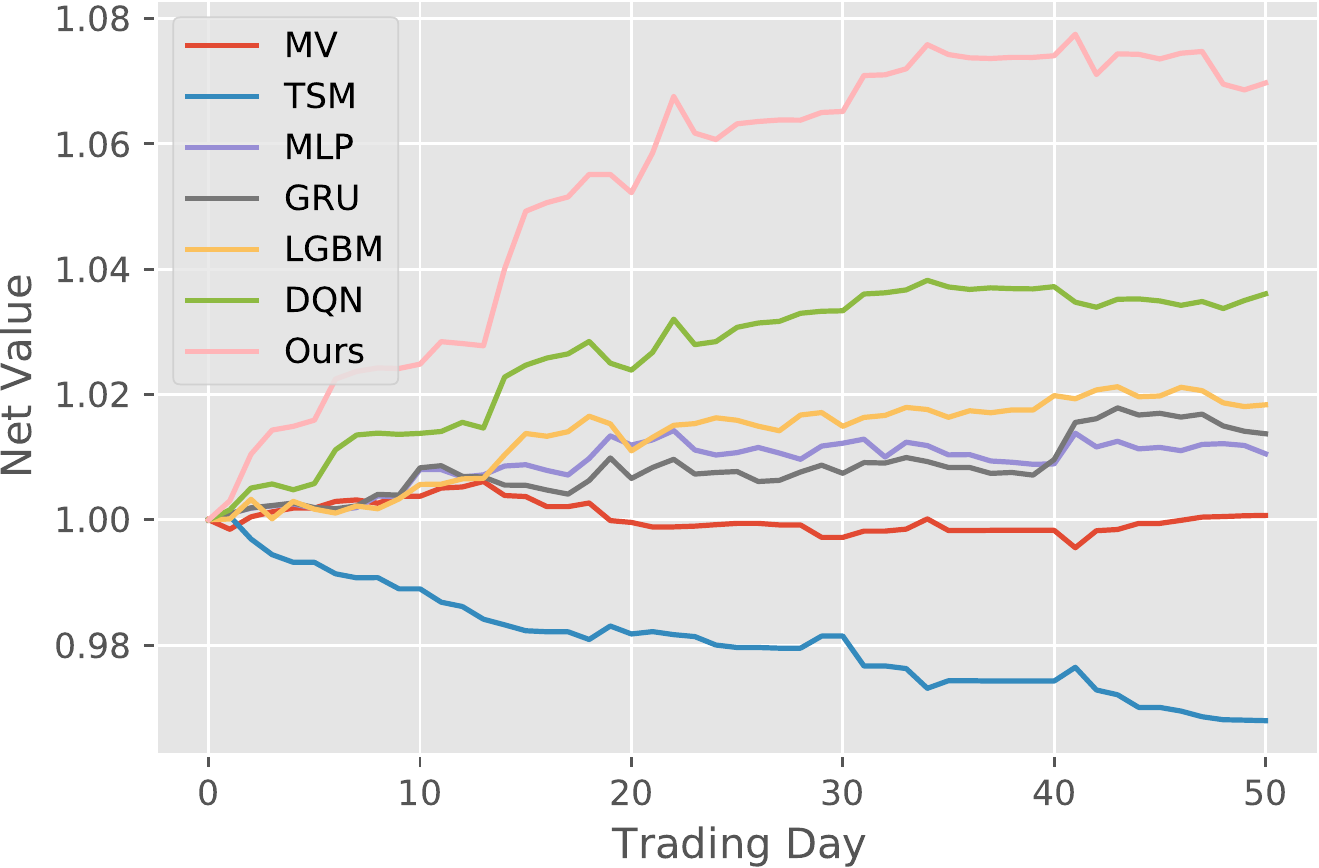}
     }
     \hfill
     \subfloat[{TF02}\label{fig:tm02}]{%
      \includegraphics[width=0.23\textwidth]{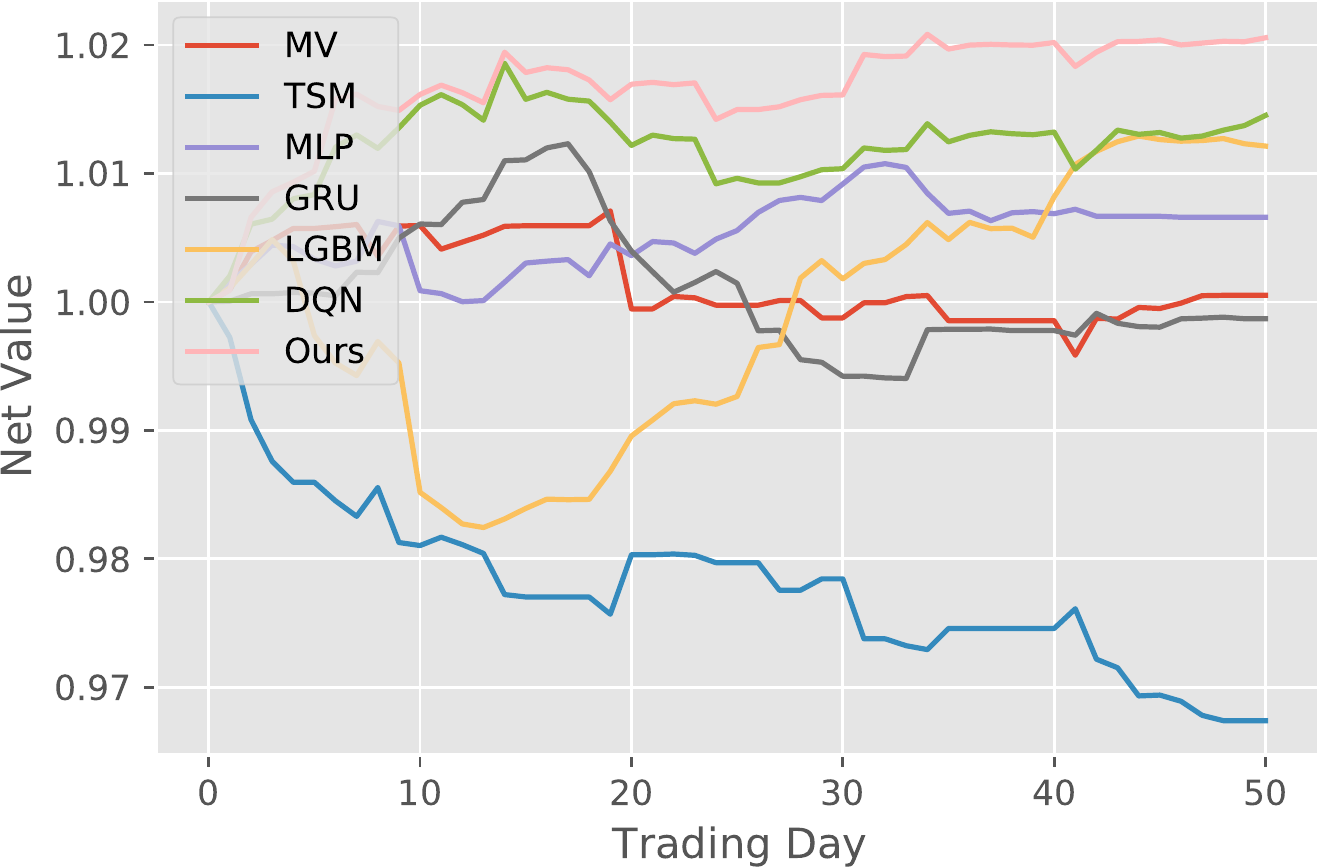}
     }
     \hfill
     \subfloat[{T01}\label{fig:antmaze0}]{%
      \includegraphics[width=0.23\textwidth]{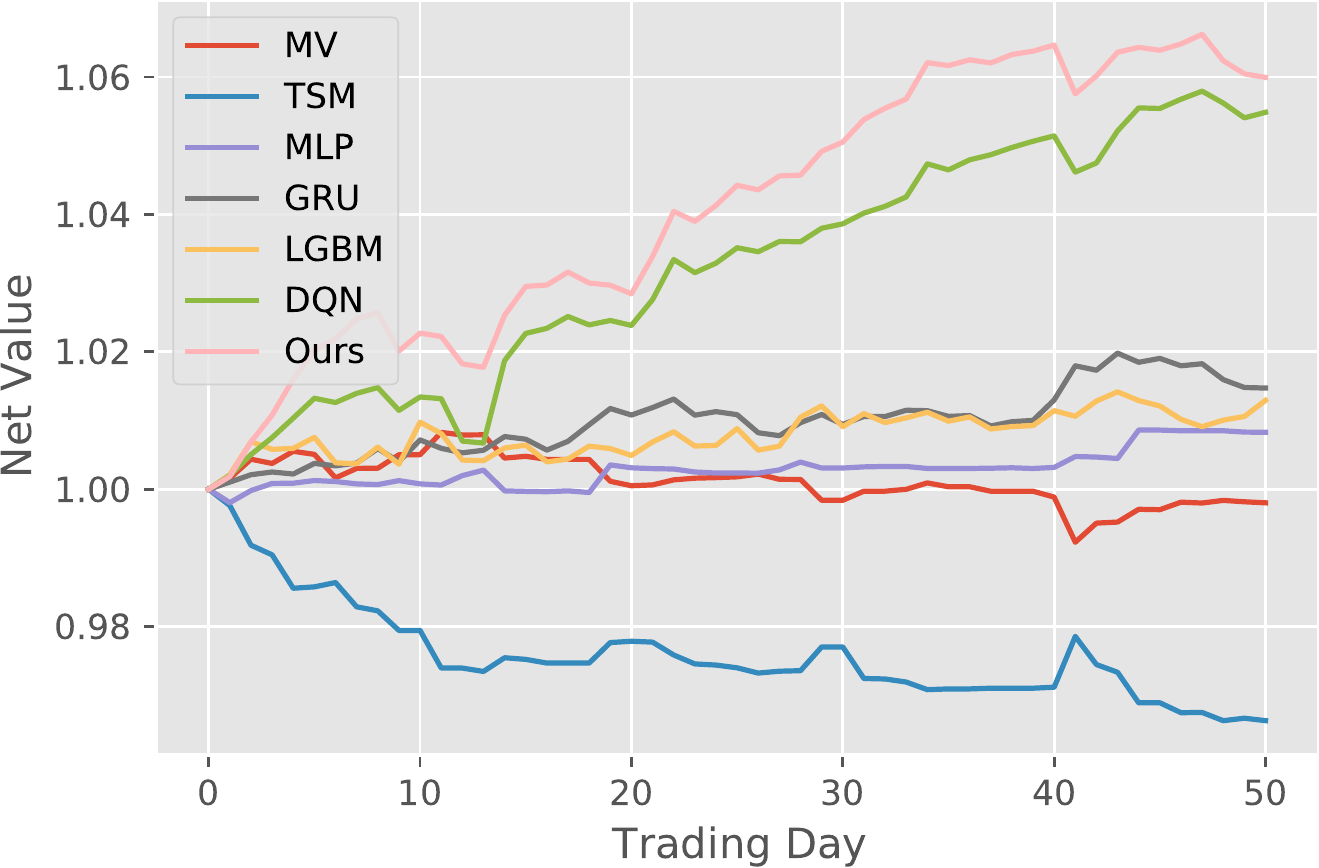}
     }
     \hfill
     \subfloat[{T02}\label{subfig-1:dummy}]{%
      \includegraphics[width=0.23\textwidth]{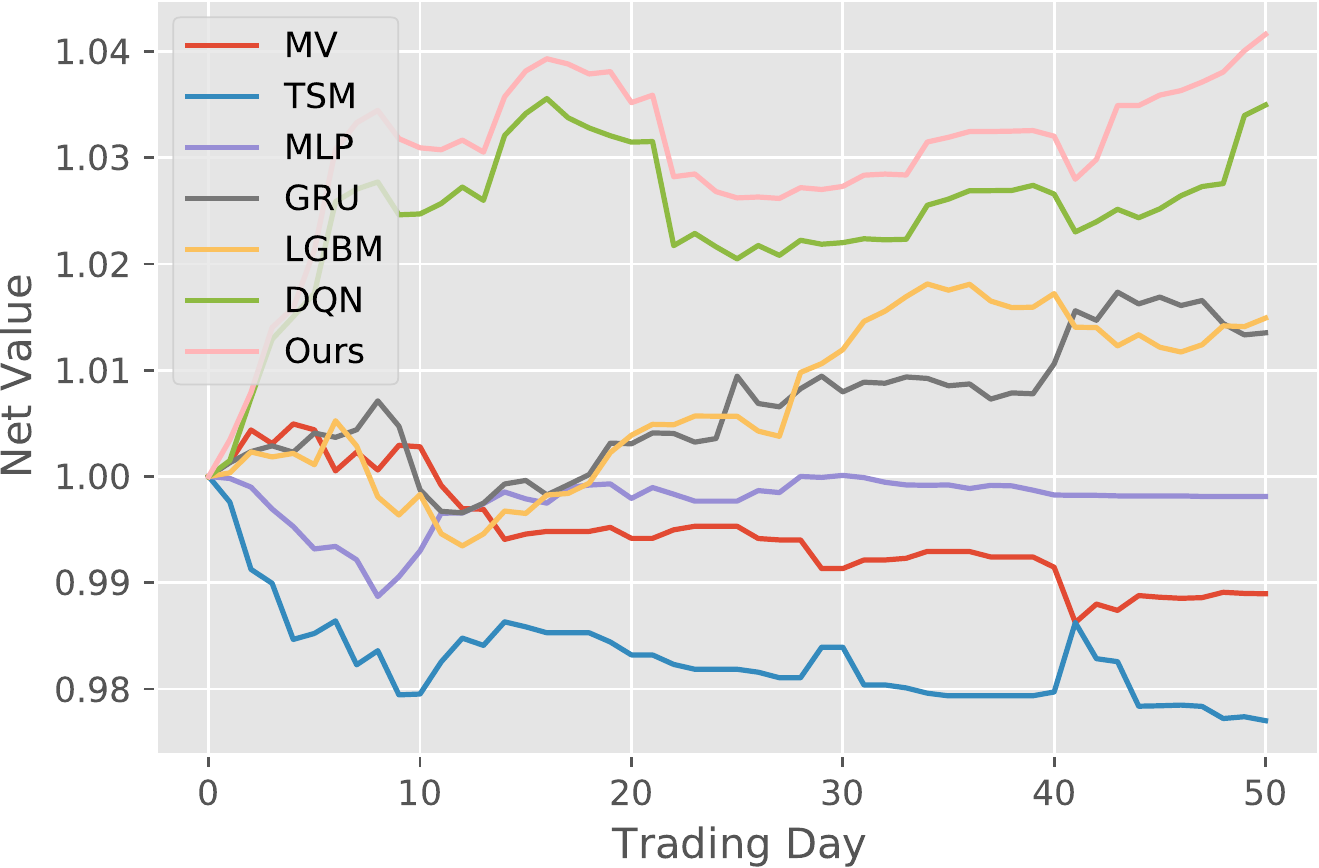}
     }
     \caption{Trading day vs. net value curve of different baselines and DeepScalper on stock index (IC and IF) and treasury bond (TF01, TF02, T01, T02) datasets. DeepScalper achieves the highest profit in all six financial assets.}
     \label{fig:net_value}
\end{figure}
\section{Results and Analysis}
\subsection{Profitability Comparison with Baselines}

We compare DeepScalper with 9 state-of-the-art baselines in terms of four financial metrics in Table \ref{fig:performance}. We observe that DeepScalper consistently generates significantly \((p<0.01)\) higher performance than all baselines on 7 out of 8 metrics across the two datasets. In the stock index dataset, DeepScalper performs best on all four metrics. Specifically, it outperforms the second best by 30.80\%, 33.33\%, 21.42\% and 7.50\% in terms of TR, SR, CR and SoR, respectively. As for the treasury bond dataset, DeepScalper outperforms the second best by 14.87\%, 7.47\%, 30.94\% in terms of TR, SR and CR. For SoR, DS-NA performs slightly better than DS (2\% without statistical significance). One possible reason is that volatility prediction auxiliary task is not directly relevant to control downside return variance. 

Furthermore, we show the trading day vs. net value trading days of the test period for each financial future from the two datasets in Figure \ref{fig:net_value}. We intentionally exclude BAH, DS-NH and DS-NA to make the figure easy to follow. For traditional methods, we find that MV achieves decent performance for most financial futures. In comparison, TSM's performance is much worse. One possible reason for TSM's failure is that there is no evident momentum effect within the market for intraday trading. For deep learning models, the overall performance of GRU is better than that of MLP due to its ability to learn the temporal dependency of indicators. As for LGBM, it achieves slightly better performance than deep learning models. The average performance of RL methods is the best.

\begin{table}[th!]
\centering
\resizebox{0.99\columnwidth}{!}{
\begin{tabular}{cccc|ll}
\hline 
Macro & Micro & Hindsight & Volatility & TR(\%)$\uparrow$ & SR$\uparrow$ \\
\hline

\(\surd\) & & & & 3.45 & 4.42 \\ 
 & \(\surd\) & & & 3.47 & 4.43 \\
\hline
\(\surd\) & \(\surd\) & & & 3.62 \textcolor[RGB]{0,162,96}{(+0.15)} & 4.81 \textcolor[RGB]{0,162,96}{(+0.38)} \\
\(\surd\) & \(\surd\) & & \(\surd\) & 4.05 \textcolor[RGB]{0,162,96}{(+0.58)} & 5.03 \textcolor[RGB]{0,162,96}{(+0.60)} \\
\(\surd\) & \(\surd\) & \(\surd\) & & 5.36 \textcolor[RGB]{0,162,96}{(+1.89)} & 5.72 \textcolor[RGB]{0,162,96}{(+1.29)} \\
\(\surd\) & \(\surd\) & \(\surd\) & \(\surd\) & \textbf{6.97} \textcolor[RGB]{0,162,96}{(+3.50)} & \textbf{6.10} \textcolor[RGB]{0,162,96}{(+1.67)} \\ \hline

\end{tabular}}
\caption{Ablation studies over different DeepScalper components. \(\surd\) indicates adding the component to DeepScalper.}\label{tab:ablation}
\end{table}

\subsection{Model Component Ablation Study}
We conduct comprehensive ablation studies on DeepScalper's investment profitability benefits from each of its components in Table \ref{tab:ablation}. First, we observe that the encoder-decoder architecture can learn better multi-modality market embedding than agents trained with other macro-level or micro-level market information, which leads to 0.15\% and 0.38 improvement of TR and SR, respectively. Next, we find that adding the volatility prediction auxiliary task into DeepScalper can further improve performance, indicating that taking risk into consideration can lead to robust market understanding. In addition, we observe that the hindsight bonus can significantly improve DeepScalper's ability for the evaluation of trading decisions and further enhance profitability. Finally, we add all these components into DeepScalper and achieve the best performance in terms of TR and SR. Comprehensive ablation studies demonstrate: 1) each individual component in DeepScalper is effective; 2) these components are largely orthogonal and can be fruitfully integrated to further improve performance.


\begin{figure}[ht]
  \centering
     \subfloat[{TF02}\label{fig:hindsight_h}]{%
      \includegraphics[width=0.23\textwidth]{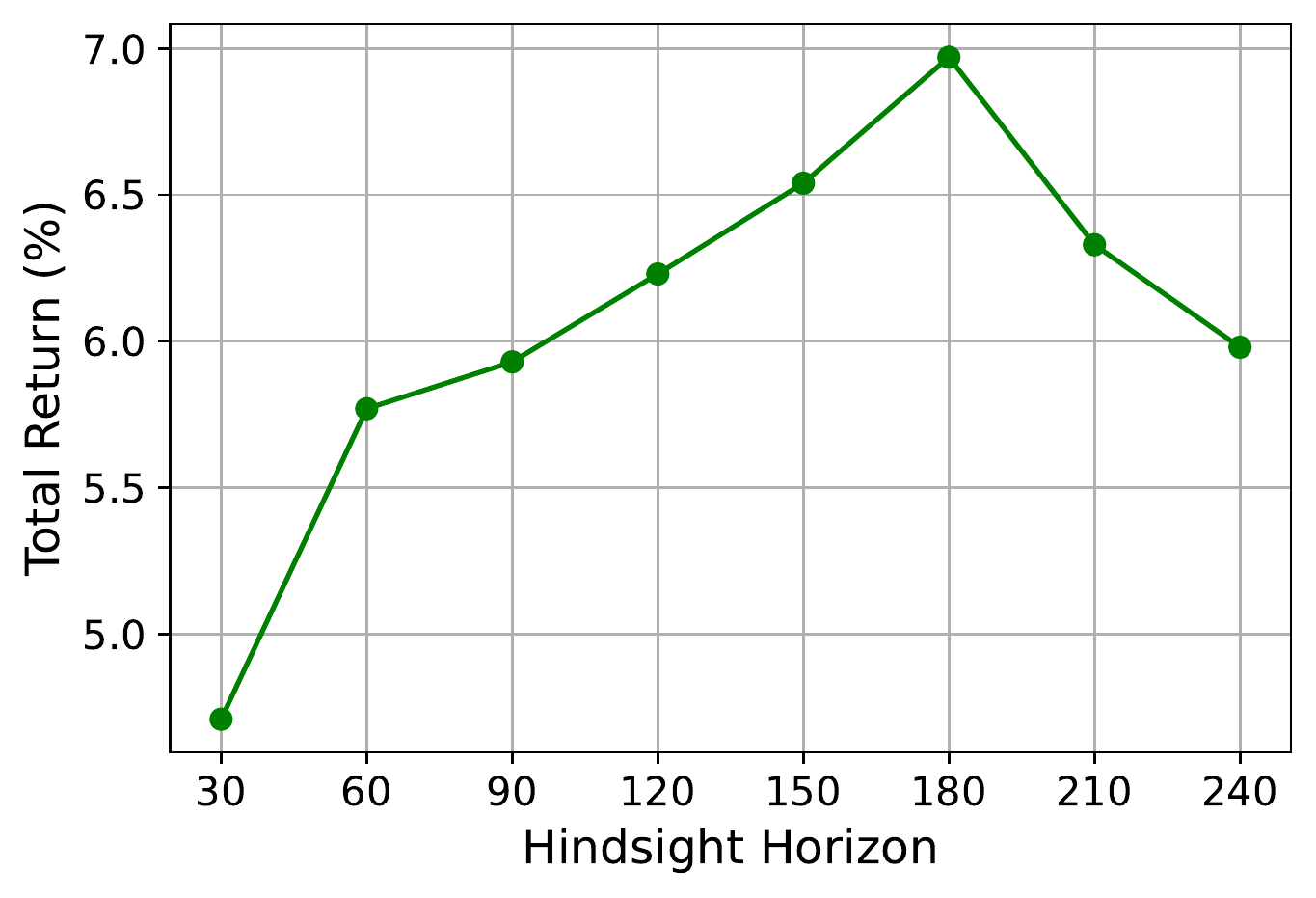}
     }
     \hfill
     \subfloat[{T02}\label{fig:hindsight_w}]{%
      \includegraphics[width=0.23\textwidth]{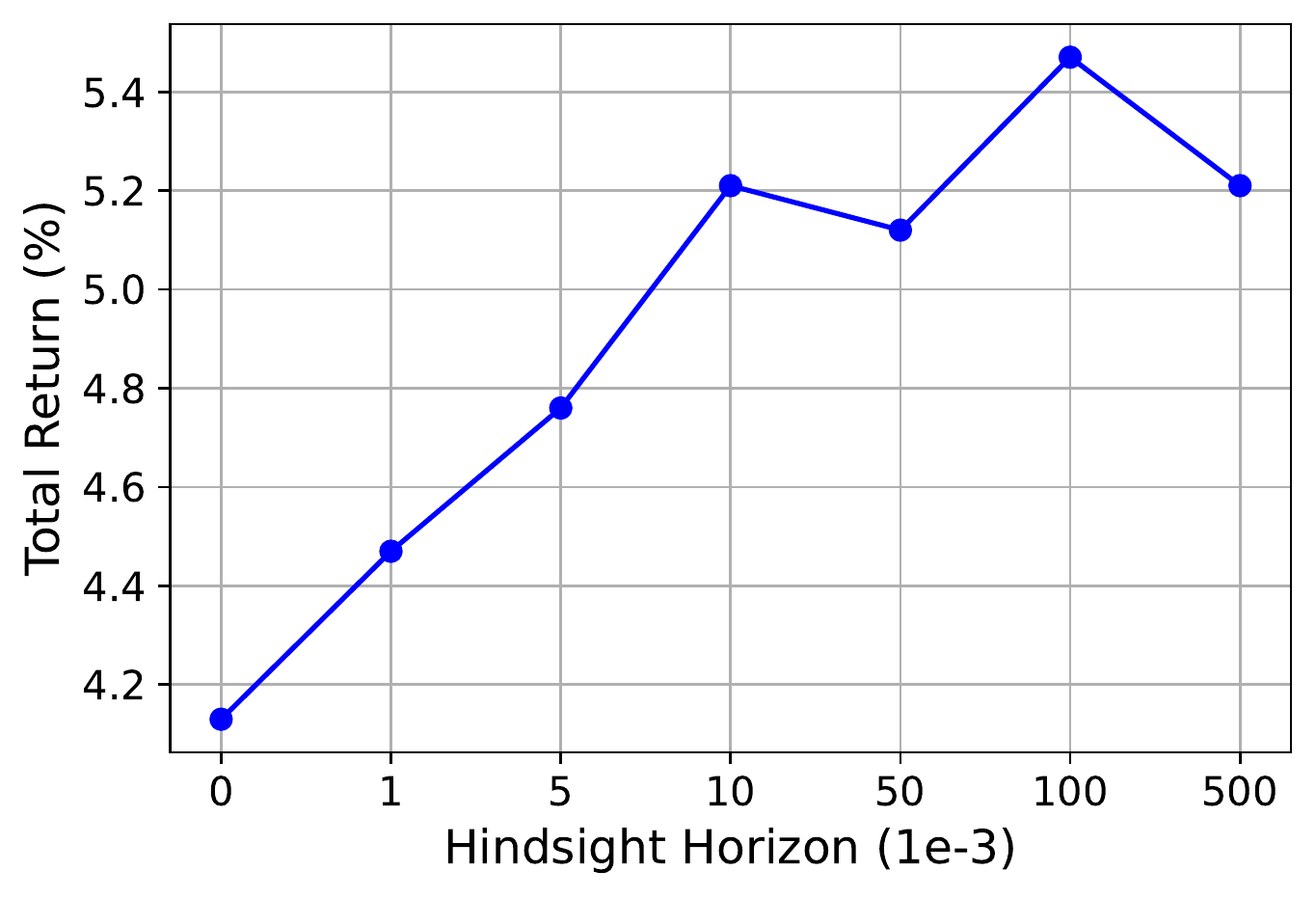}
     }
     \caption{Hyperparameter sensitivity of hindsight bonus: (a) effect of importance (b) effect of horizon}
     \label{fig:hindsight_bonus_ablation}
\end{figure}

\begin{figure}[!bht]
  \centering
     \subfloat[{May 26 (DS-NH)}\label{fig:ds-nh}]{%
      \includegraphics[width=0.23\textwidth]{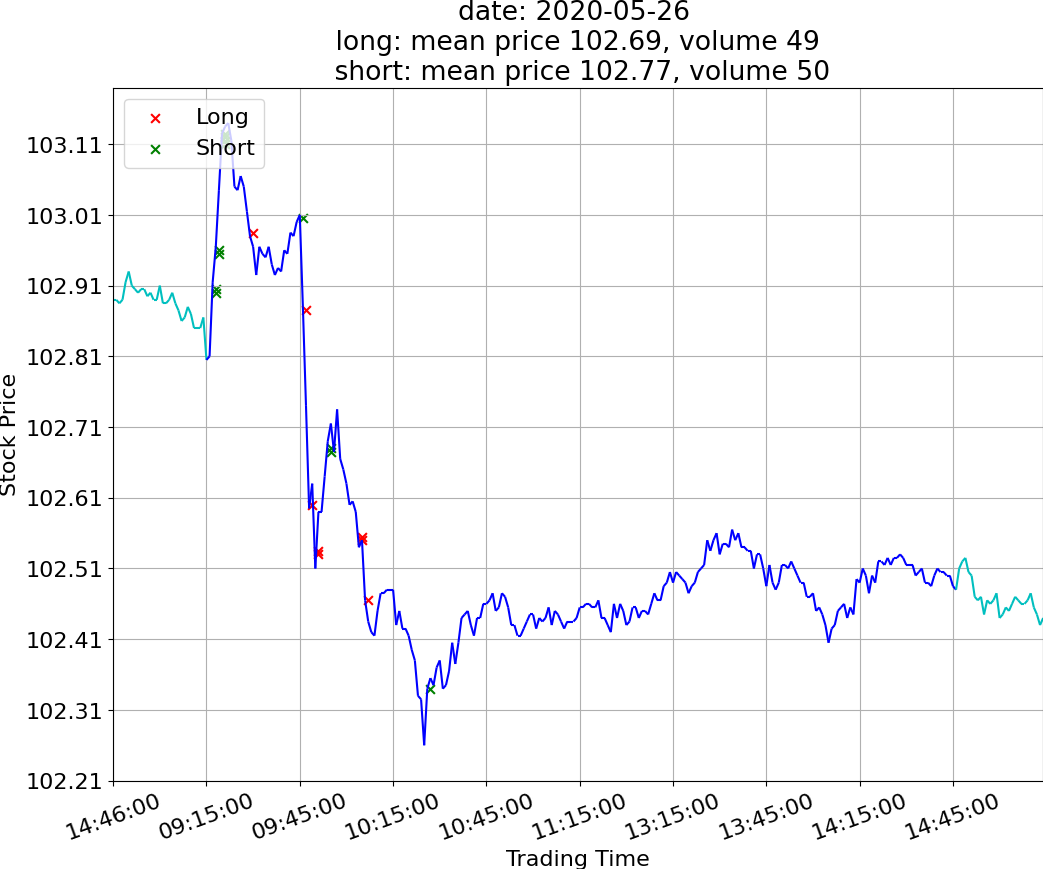}
     }
      \hfill
     \subfloat[{May 26 (DS)}\label{fig:ds}]{%
      \includegraphics[width=0.23\textwidth]{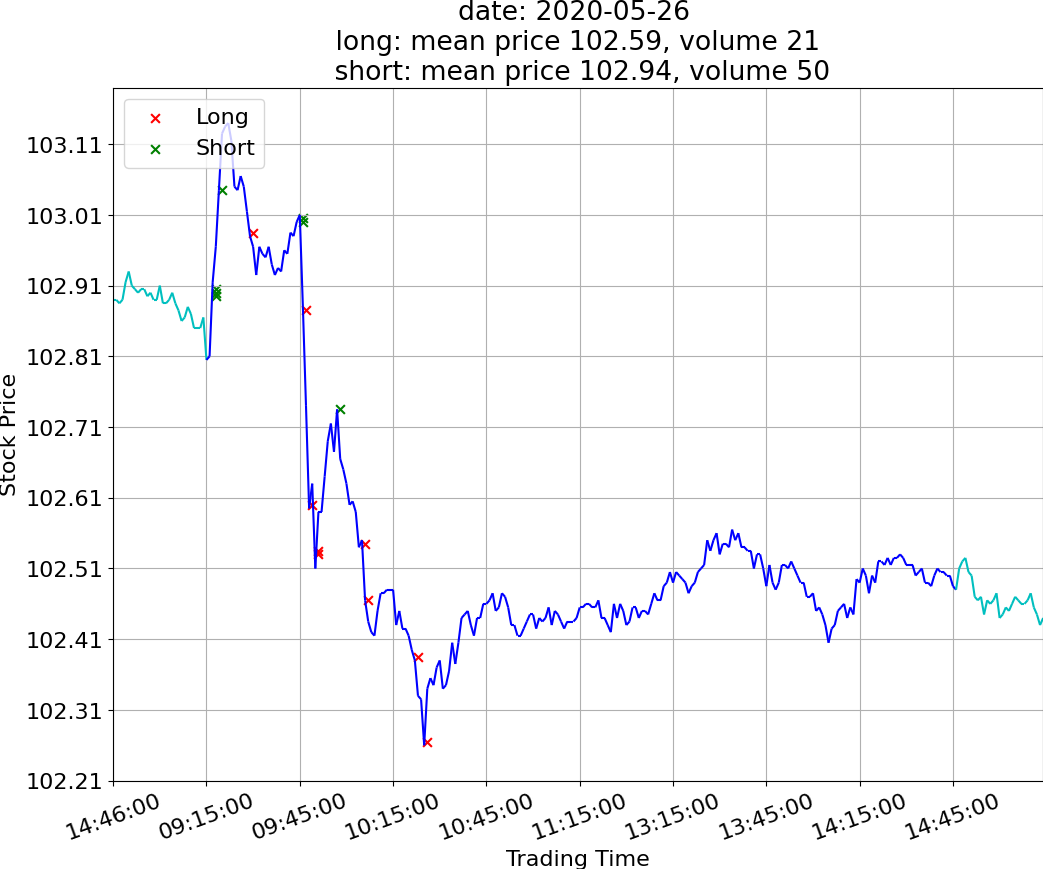}
     }
     \hfill

     \caption{Trading behavior comparison of DS-NH and DS to show the effectiveness of the hindsight bonus}
     \label{fig:trading_behavior}
\end{figure}

\subsection{Effectiveness of Hindsight Bonus}
We analyze the effectiveness of the hindsight bonus from two perspectives. First, we explore the impact of the hindsight bonus horizon and weight. As shown in Figure \ref{fig:hindsight_h}, with the increase of \(w\), the agent tends to trade with a long-term horizon and achieves a higher profit. DeepScalper with \(w=0.1\) achieves the highest profit. Figure \ref{fig:hindsight_w} shows the impact of hindsight horizon \(h\) on DeepScalper's performance. We observe that DeepScalper's total return gradually increases by moving \(h\) from 30 to 180 and decreases when \(h>180\). 

Moreover, we compare the detailed trading behaviors of agents trained with and without hindsight bonus on a trading day with the decreasing trend in Figure \ref{fig:trading_behavior}. The general good intraday trading strategy for that day is to short at the start of the day and long at the end of the day. We find that the agent trained without the hindsight bonus (Figure \ref{fig:ds-nh}) performs well in capturing local trading opportunities and overlooks the long-term trend of the entire trading day. In comparison, an agent trained with the hindsight bonus (Figure \ref{fig:ds}) trades a large volume of short actions at the beginning of the trading day, indicating that it is aware of the decreasing trend in advance. This kind of trading action is smart, since it captures the big price gap of the overall trend and somehow ignores the local gain or loss.

\subsection{Effectiveness of Risk-aware Auxiliary Task}
\begin{figure}[t]
\begin{center}
\includegraphics[width=0.7\columnwidth]{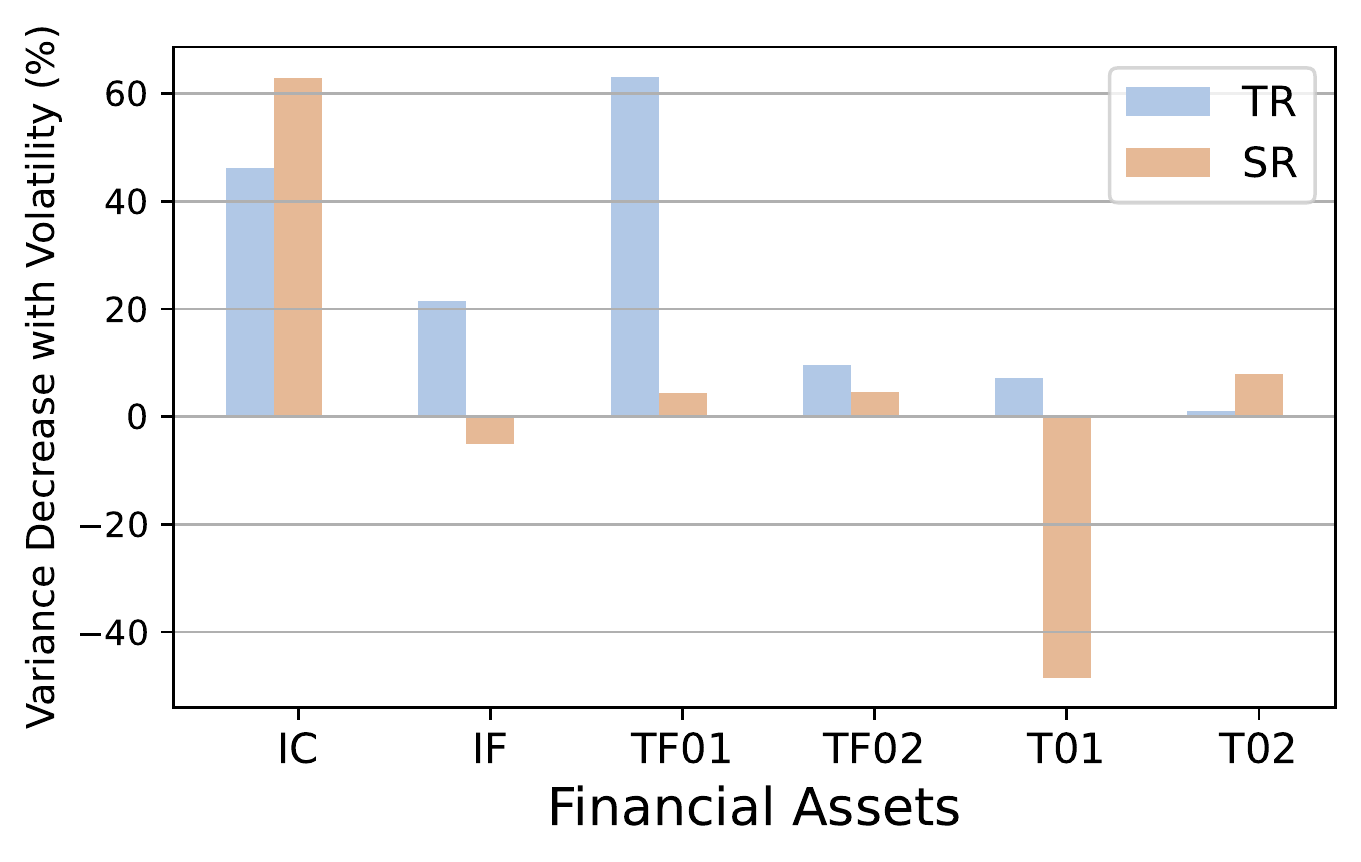}
\caption{Effect of the auxiliary task on performance variance (\(>\)0 means RL agents trained with the risk-aware auxiliary task get lower standard deviation)}
\label{fig:variance_aux}
\end{center}
\end{figure}

\begin{figure}[t]
\begin{center}
\includegraphics[width=0.7\columnwidth]{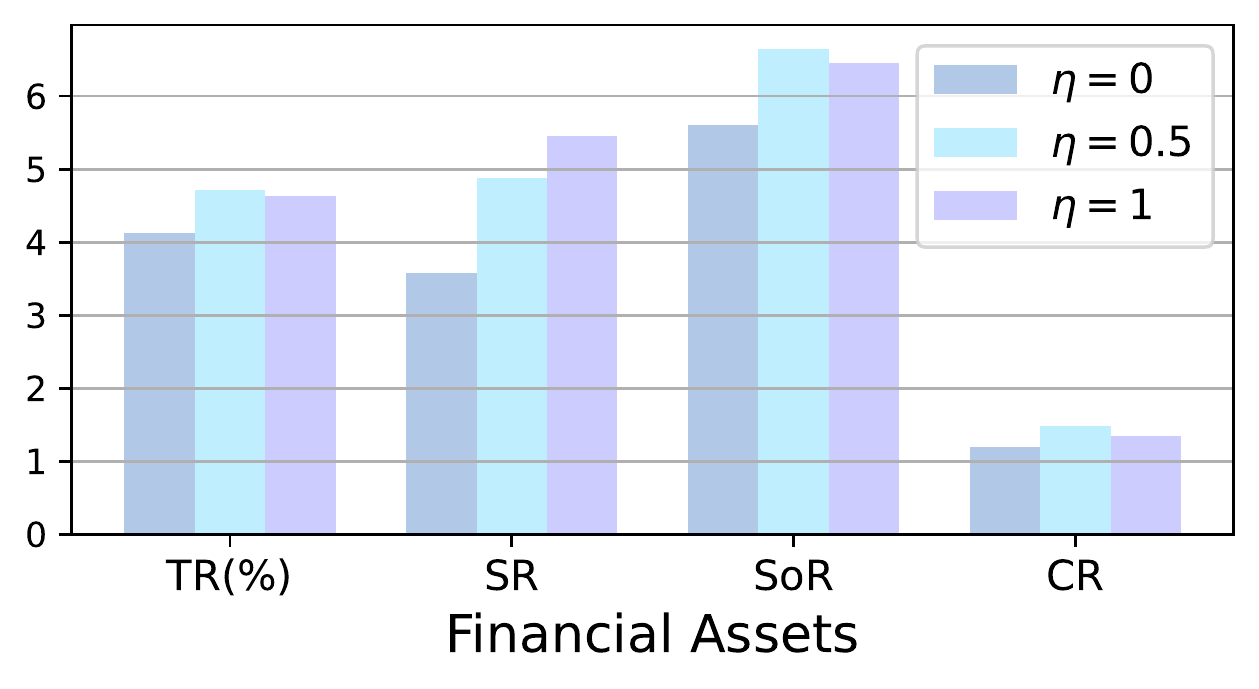}
\caption{Sensitivity to relative importance \(\eta\) in terms of TR, SR, SoR and CR}
\label{fig:aux_importance}
\end{center}
\end{figure}

Since the financial market is noisy and the RL training process is unstable, the performance variance among different random seeds is a major concern of RL-based trading algorithms. Intuitively, taking market risk into account can help the RL agent behave more stable with lower performance variance. We run experiments 5 times with different random seeds and report the relative variance relationship between RL agents trained with/without the risk-aware auxiliary task in Figure \ref{fig:variance_aux}. We find that RL agents trained with the risk-aware auxiliary task achieve a lower TR variance in all six financial assets and a lower SR variance in 67\% of financial assets. Furthermore, we test the impact of auxiliary task importance \(\eta\) on DeepScalper's performance. Naturally, the volatility value scale is smaller than return, which makes \(\eta=1\) a decent option to start. In practice, we test \(\eta\in[0, 0.5, 1]\) and find the improvement of the auxiliary task is robust to different importance weights as shown in Figure \ref{fig:aux_importance}.

\subsection{Generalization Ability}

\begin{figure}[th!]
  \centering
     \subfloat[{TF02}\label{fig:tfm02}]{%
      \includegraphics[width=0.23\textwidth]{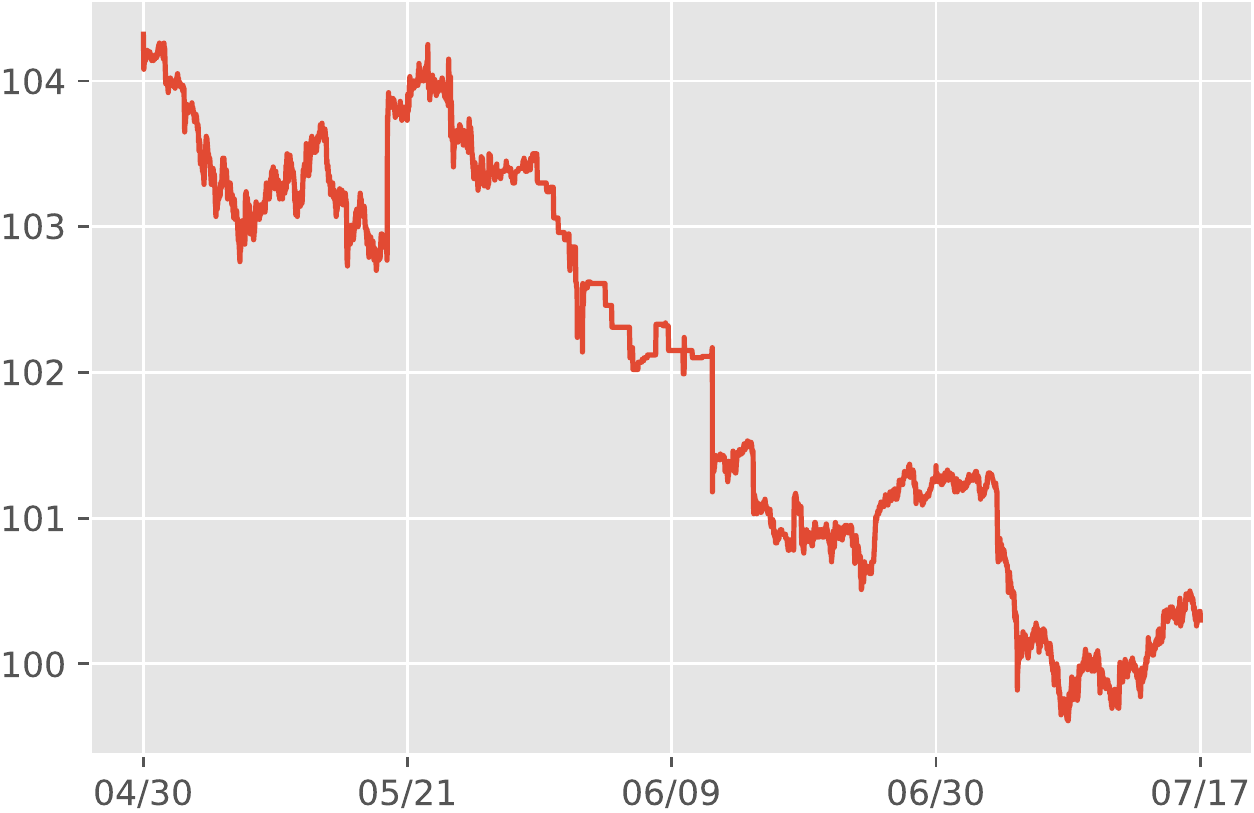}
     }
     \hfill
     \subfloat[{T02}\label{fig:tm02}]{%
      \includegraphics[width=0.23\textwidth]{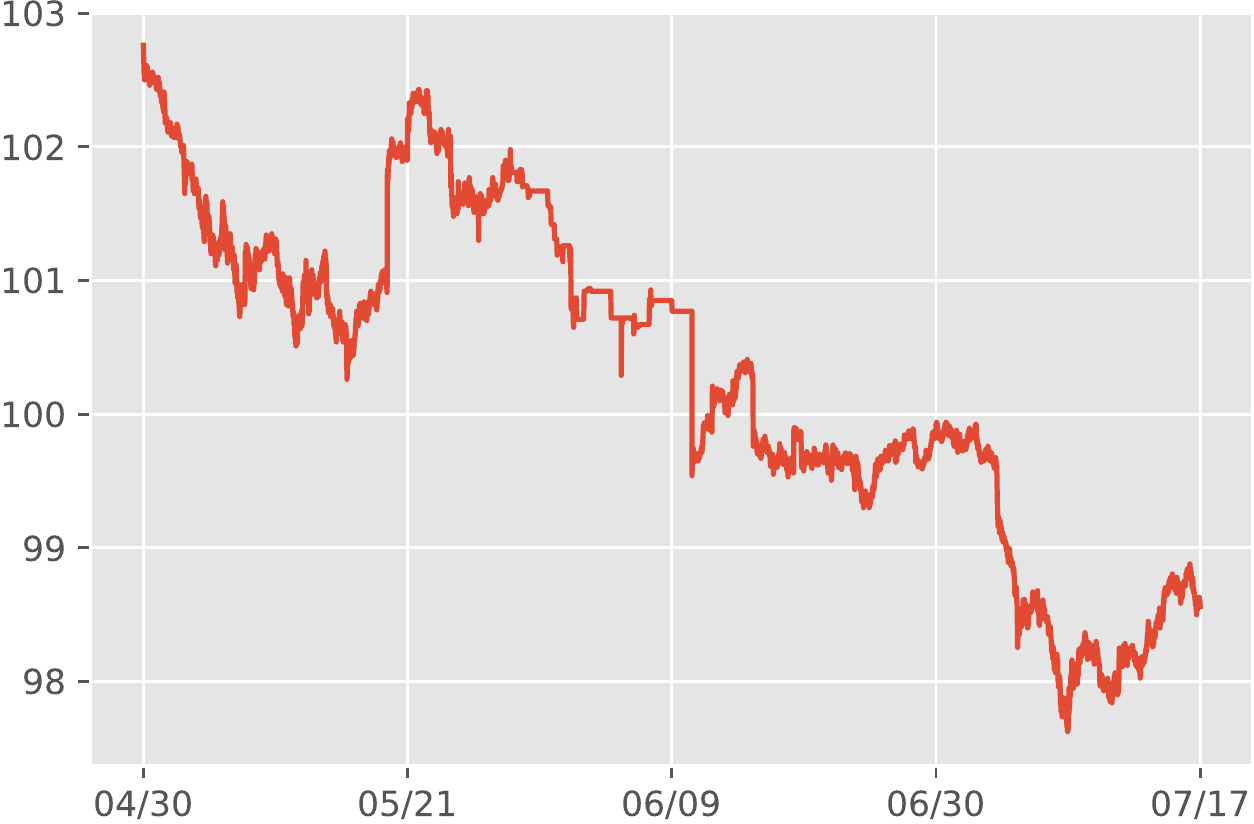}
     }
     \caption{Price curve of TF02 and T02}
     \label{fig:price_curve}
\end{figure}

\begin{figure}[th!]
  \centering
     \subfloat[{MV}\label{fig:mv}]{%
      \includegraphics[width=0.23\textwidth]{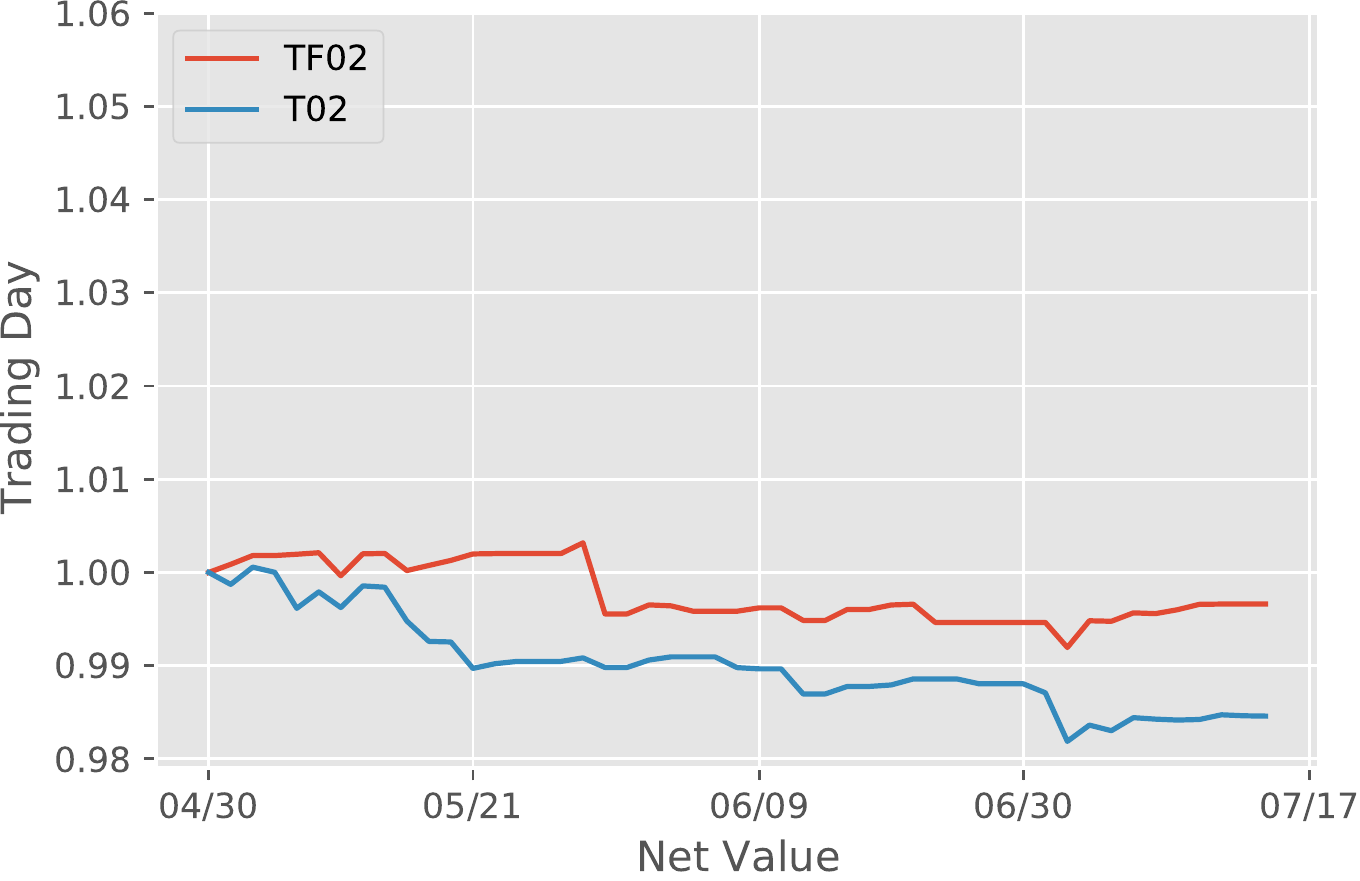}
     }
     \hfill
     \subfloat[{GRU}\label{fig:ours}]{%
      \includegraphics[width=0.23\textwidth]{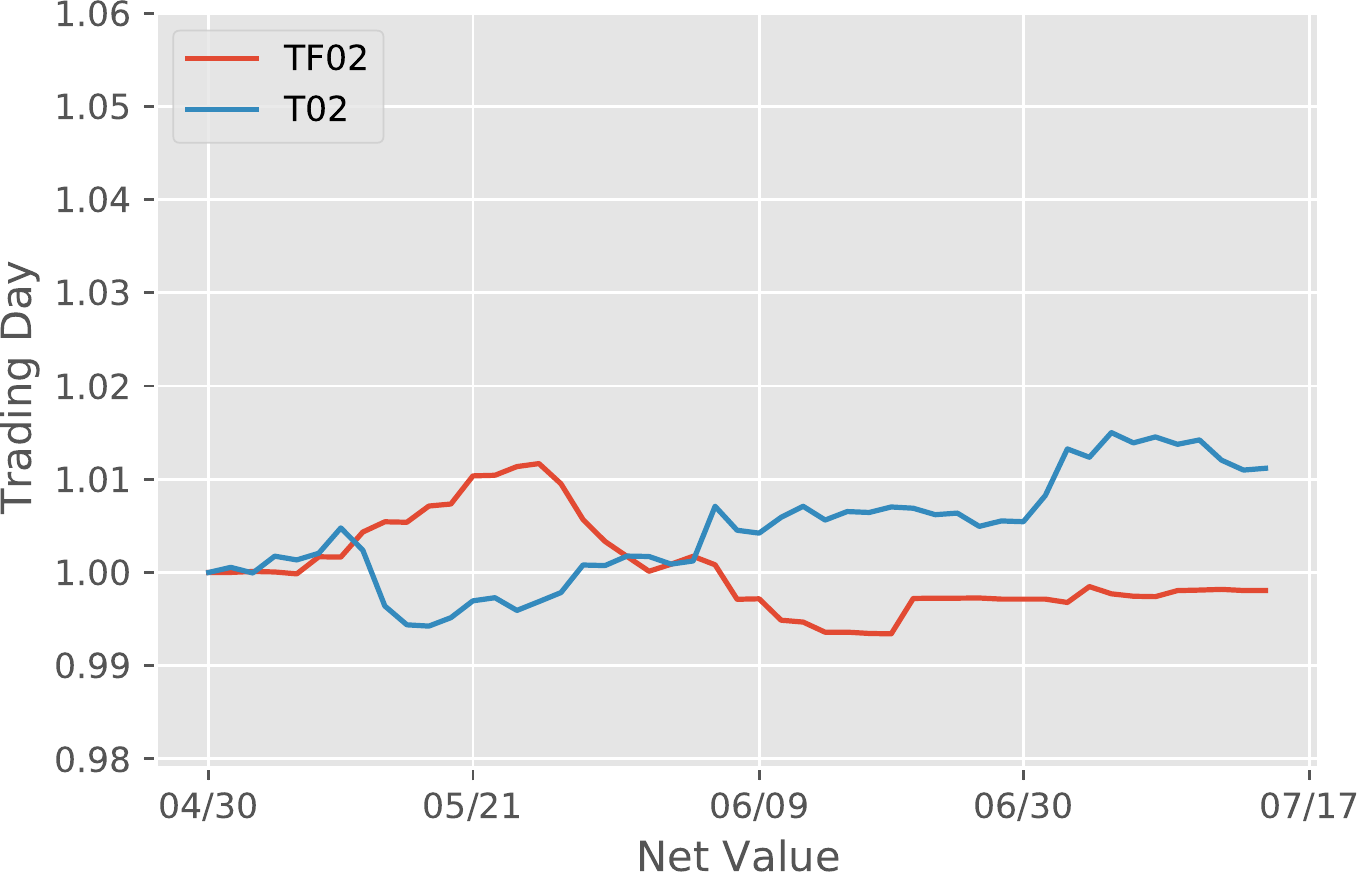}
     }
     \hfill
     \subfloat[{LGBM}\label{fig:mlp}]{%
      \includegraphics[width=0.23\textwidth]{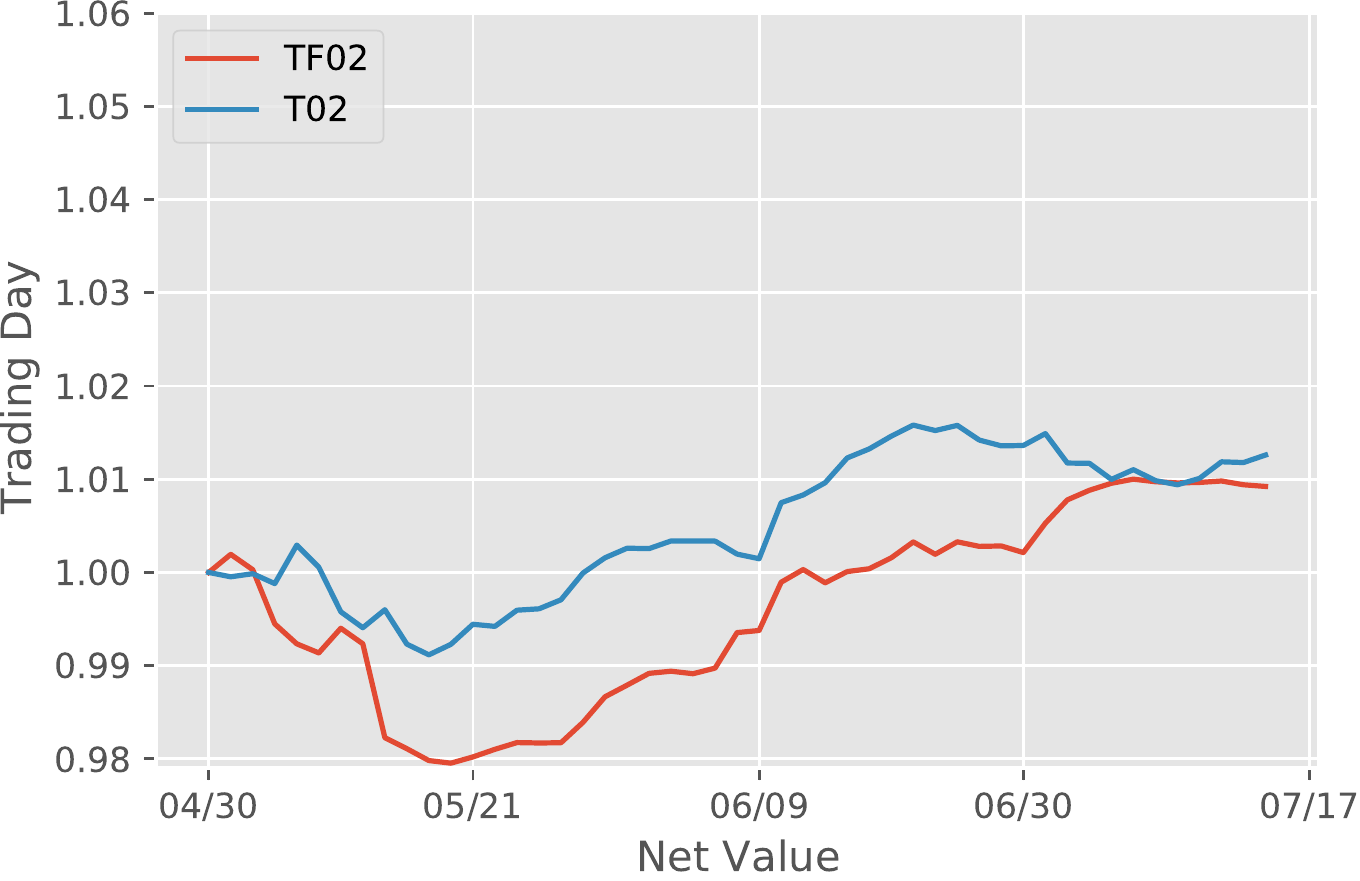}
     }
     \hfill
     \subfloat[{DeepScalper}\label{fig:mlp}]{%
      \includegraphics[width=0.23\textwidth]{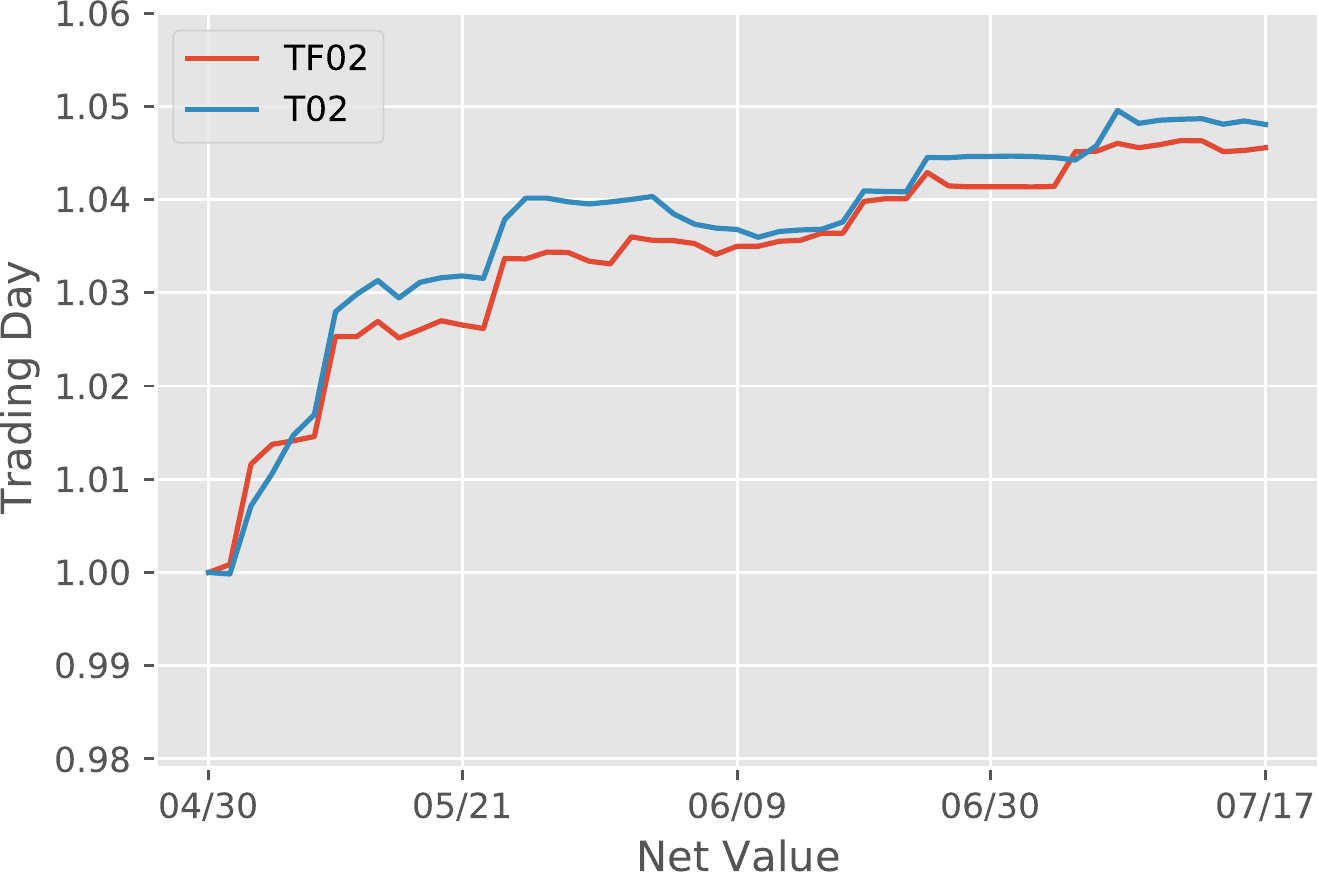}
     }
     \caption{Net value curve of MV, GRU, LGBM and DeepScalper on TF02 and T02 to show the generalization ability.}
     \label{fig:generalization}
\end{figure}

We further test the generalization ability of our framework among different financial futures (TF02 and T02). In Figure \ref{fig:price_curve}, it is clear that the price trend of TF02 and T02 is similar. We assume similar price curves shares and similar trading patterns. Then, we train DeepScalper using the TF02 training set and test it on the test set of both TF02 and T02. We compare the performance of MV, GRU, LGBM, and DeepScalper in Figure \ref{fig:generalization}. The red and blue lines represent the performance on TF02 and T02, respectively. We observe that the performance of MV, GRU, and LGBM among these two assets is quite different, demonstrating that they have poor generalization ability on our task. One possible reason is that the trading signal of MV, GRU and LGBM involves heuristic rules or threshold. There will be some potential trading opportunities that are close to meet the trading rules or threshold, but MV, GRU, and LGBM will lose the opportunities. At the same time, our DeepScalper achieves robust performance on both TF02 and T02 as shown in Figure \ref{fig:generalization} (d), although it has never seen T02 data before. All these experiments demonstrate that DeepScalper can learn a robust representation of the market and achieve good generalization ability.

\section{Conclusion}
In this article, we focus on intraday trading and propose DeepScalper to mimic the workflow of professional intraday traders. First, we apply the dueling Q-network with action branching to efficiently train intraday RL agents. Then, we design a novel reward function with a hindsight bonus to encourage a long-term horizon to capture the overall price trend. In addition, we design an encoder-decoder architecture to learn robust market embedding by incorporating both micro-level and macro-level market information. Finally, we propose volatility prediction as an auxiliary task to help agents be aware of market risk while maximizing profit. Extensive experiments on two stock index futures and four treasury bond futures demonstrate that DeepScalper significantly outperforms many advanced methods.


\bibliographystyle{ACM-Reference-Format}
\balance
\bibliography{reference}

\end{document}